\documentclass[english]{article}
\usepackage[T1]{fontenc}
\usepackage[utf8]{luainputenc}
\usepackage{geometry}
\usepackage{babel}
\usepackage{float}
\usepackage{textcomp}
\usepackage{tipa}
\usepackage{amsmath}
\usepackage{amsthm}
\usepackage{graphicx}
\usepackage{setspace}
\usepackage[unicode=true,pdfusetitle,
 bookmarks=true,bookmarksnumbered=false,bookmarksopen=false,
 breaklinks=false,pdfborder={0 0 1},backref=false,colorlinks=false]
 {hyperref}
\begin{document}
\begin{center}
\textbf{\LARGE{}The Concentration Risk Indicator: Raising the Bar
for Financial Stability and Portfolio Performance Measurement}{\LARGE\par}
\par\end{center}

\begin{center}
\textbf{\large{}Ravi Kashyap (ravi.kashyap@stern.nyu.edu)}\footnote{\begin{doublespace}
Numerous seminar participants, particularly at a few meetings of the
econometric society and various finance organizations, provided suggestions
to improve the paper. The following individuals have been a constant
source of inputs and encouragement: Dr. Yong Wang, Dr. Isabel Yan,
Dr. Vikas Kakkar, Dr. Fred Kwan, Dr. Costel Daniel Andonie, Dr. Guangwu
Liu, Dr. Jeff Hong, Dr. Humphrey Tung and Dr. Xu Han at the City University
of Hong Kong. The views and opinions expressed in this article, along
with any mistakes, are mine alone and do not necessarily reflect the
official policy or position of either of my affiliations or any other
agency.
\end{doublespace}
}
\par\end{center}

\begin{center}
\textbf{\large{}Estonian Business School / City University of Hong
Kong}{\large\par}
\par\end{center}

\begin{center}
\today
\par\end{center}

\begin{center}
Keywords: Risk Management; Concentration Risk Indicator (CRI); Financial
Stability; Portfolio Performance Measurement; Quantitative Marketing
Metrics; Asset Price; Volatility; Blockchain
\par\end{center}

\begin{center}
Journal of Economic Literature Codes: C43 Index Numbers and Aggregation;
G11 Investment Decisions; D81 Criteria for Decision-Making under Risk
and Uncertainty; B23 Econometrics, Quantitative and Mathematical Studies
\par\end{center}

\begin{center}
Mathematics Subject Classification Codes: 91G15 Financial markets;
91G10 Portfolio theory; 62M10 Time series; 91G70 Statistical methods,
risk measures; 91G45 Financial networks
\par\end{center}

\pagebreak{}
\begin{center}
\tableofcontents{}
\par\end{center}

\begin{center}
\pagebreak{}
\par\end{center}

\begin{doublespace}
\begin{center}
\listoffigures 
\par\end{center}

\begin{center}
\listoftables 
\par\end{center}
\end{doublespace}

\begin{singlespace}
\begin{center}
\pagebreak{}
\par\end{center}
\end{singlespace}
\begin{doublespace}

\section{Abstract}
\end{doublespace}

We have developed a novel risk management measure called the concentration
risk indicator (CRI). The CRI has been created to address drawbacks
with prevailing methodologies and to supplement existing methods.
Modified and adapted from the Herfindahl-Hirschman (HH) index, the
CRI can give a single numeric score that can be helpful to evaluate
the extent of risks that arise from holding concentrated portfolios.
We discuss how the CRI can become an indicator of financial stability
at any desired aggregation unit: regional, national or international
level. The concentration risk indicator can be calculated for individual
assets and for portfolios of assets as well. We show how the CRI can
be easily applied to insurance risk and to any product portfolio mix. 

The CRI - when used alongside other well known portfolio measurement
yardsticks - allows us to gauge how portfolios are performing and
to compare them to the wider set of investment opportunities. Such
an approach will give a more complete picture of how well any investment
machinery is working. When comparing two investments, the lower the
concentration risk, the better the investment from a diversification
point of view. The CRI is particularly applicable to the current facet
of the decentralized terrain, wherein the majority of the wealth is
restricted to a small number of tokens. We calculate and report the
CRI - along with other risk metrics - for individual assets and portfolios
of crypto assets using a daily data sample from January 01, 2019 until
August 10, 2022. 

The CRI is an example of developing metrics that can useful for sending
concise yet powerful messages to the relevant audience. This tactic
- which can be described as marketing the benefits of any product
or service by using concepts from multiple disciplines - of creating
new metrics goes further beyond the use of metrics to evaluate marketing
efficacy. The simplicity of our metric - and the intuitive explanations
we have provided for the CRI - makes it straightforward to properly
articulate a strong - clear and positive - signal as part of marketing
campaigns. 

The development - and implementation - of new risk management metrics
will have greater impact when a wider rigorous risk management process
has been established. We discuss several topics related to bringing
about more improved risk management across all types of institutions
and assets. 

\section{\label{sec:Portfolio-Performance-Problems}Introduction: Portfolio
Performance Problems }

The blockchain phenomenon offers numerous possibilities to transform
all aspects of human interaction (Nakamoto 2008; Di Pierro 2017; Chen
2018). Bringing Risk Parity to the DeFi Party (Fabozzi, Simonian \&
Fabozzi 2021; Zetzsche, Arner \& Buckley 2020; Jensen, von Wachter
\& Ross 2021; Werner et al., 2021; Mohan 2022; Grassi et al., 2022;
End-notes \ref{enu:Risk-parity-(or}; \ref{enu:Decentralized-finance})
has been the impetus for numerous innovations and designs described
in Kashyap (2022). To build a robust decentralized investment management
platform, coupled with rigorous risk management, we have identified
seven fundamental themes that need to be incorporated: 1) DeFi Security;
2) Rebalancing and Trade Execution; 3) Asset Weight Calculation; 4)
Risk Parity Construction; 5) Profit Sharing and Investor Protection;
6) Concentration Risk Indicator and Portfolio Performance Metrics;
and 7) Multi-chain expansion and Select Strategic Initiatives. Once
these novel techniques, are implemented in the blockchain environment,
it will create an unparalleled platform for wealth generation accessible
by anyone. 

In this article we will discuss a new metric we have developed - termed
the concentration risk indicator (CRI) - that will allow us to gauge
how portfolios are performing and compare them to the wider set of
crypto investment opportunities. This metric is focused on the current
facet of the decentralized terrain, wherein the majority of the wealth
is restricted to a small number of tokens. Our new measure, when supplemented
with other well known portfolio measurement benchmarks, will give
a complete picture of how well any investment machinery is working.
Surely, portfolio performance measurement is an active and wonderfully
researched area with many issues being highlighted and interesting
measures being developed (Eccles 1991; Chen \& Knez 1996; Murthi,
Choi \& Desai 1997; Goetzmann et al. , 2007; Marhfor 2016).

The CRI is a novel risk management measure created to address drawbacks
with prevailing methodologies and to supplement existing methods.
The CRI can give a single numeric score that can be helpful to evaluate
the extent of risks that arise from holding concentrated portfolios.
When comparing two investments, the lower the concentration risk,
the better the investment from a diversification point of view. To
illustrate how this measure works in practice, we calculate and report
the CRI - along with other risk metrics - for individual assets and
portfolios of crypto assets using a daily data sample from January
01, 2019 until August 10, 2022. 

We discuss how the CRI can become an indicator of financial stability
at any desired aggregation unit: regional, national or international
level. The concentration risk indicator can be calculated for individual
assets and for portfolios of assets as well - across any asset class
and even combinations of asset classes. It has applications outside
financial portfolio management by looking at the concentration of
diverse portfolios of products, services, insurance contracts and
other entities that require risk mitigation.

The CRI is an example of developing metrics that can useful for sending
concise yet powerful messages to the relevant audience. This tactic
- which can be described as marketing the benefits of any product
or service by using concepts from multiple disciplines - of creating
new metrics goes further beyond the use of metrics to evaluate marketing
efficacy (Lehmann 2004; Bendoly et al., 2007; Seggie et al., 2007;
Ambler 2008; Edeling et al., 2021; Csikósová et al., 2016). The simplicity
of our metric - and the intuitive explanations we have provided for
the CRI - makes it straightforward to properly articulate a strong
- clear and positive - signal as part of marketing campaigns. 

The blockchain landscape has many individuals who invested early in
projects such as Bitcoin and Ethereum, when they were up and coming
prospects (Bohr \& Bashir 2014; Wolfson 2015; Chowdhury 2016; England
\& Fratrik 2018; Boreiko \& Risteski 2021; Fahlenbrach \& Frattaroli
2021; Noda 2021; Ante et al., 2022). These initial holdings have grown
significantly to become fairly large positions. From a portfolio perspective,
the wealth of these early adopters is heavily concentrated in a few
names. This is also the very nature of the crypto markets, where bitcoin
and ethereum command more than 60\% of the total market capitalization
(Wu, Wheatley \& Sornette 2018; Akbulaev, Mammadov \& Hemdullayeva
2020; Chang \& Shi 2020; Allen, Fatás \& di Mauro 2022; Urquhart 2022;
End-note \ref{enu:Bitcoin-(BTC)-accounted}; \ref{enu:Market-capitalization,-sometimes}).

The number of tokens listed now on major crypto data providers, such
as CoinMarketCap, is around 19,500+ as of May-25-2022 (Roosenboom
et al., 2020; End-note \ref{enu:CoinMarketCap}). This figure has
more than doubled within the last one year. With a trend where several
tokens appear - and an equal or greater number disappear - choosing
the right investments is an arduous task. Proper due diligence and
research procedures need to be utilized for forming portfolios. Having
the right selection methodology is crucial and, once a selection is
made, evaluating the corresponding performance is equally important
(Brown, Fraser \& Liang 2008; Travers 2011; Guenther, Hienerth \&
Riar 2015; Cumming, Johan \& Zhang 2019; Pástor 2000; Liu et al.,
2021; Rzepczynski \& Black 2021). To address drawbacks with prevailing
methodologies - and to supplement existing methods - we had to come
up with the CRI.

The development - and implementation - of new risk management metrics
will have greater impact when a wider rigorous risk management process
has been established. We discuss several topics related to bringing
about more improved risk management across all types of institutions
and assets. We point out the primary riddle of risk management, which
is that the life of a risk manager is very similar to someone doing
a slow walk on a tight rope. 

We provide a set of directives - the five pillars risk mitigation
guidelines - aimed at brevity so that it is readily apparent what
actions to perform, and implement, without much confusion. It is generally
a troubling scenario wherein someone tries to manage something that
they do not fully understand. Risk, as we know it, requires a lot
more in-depth studies to be performed before we are in a position
to more fully appreciate the etiology of risk. With additional research
and understanding, we will be in a better place to try and manage
risk more throughly. Our present attempts are better referred to as
Risk Mitigation.

At present, blockchain investing is seen as too risky due to the huge
swings that crypto portfolios go through on a regular basis. By creating
superior risk managed portfolios the entire decentralized finance
space will become more appealing to investors, spurring the growth
of decentralized innovations. The next chapters of the extremely promising
blockchain technological saga have to be written on the foundation
of better risk principles.

\subsection{\label{subsec:Outline-of-the}Outline of the Sections Arranged Inline}

Section (\ref{sec:Portfolio-Performance-Problems}), which we have
already seen, provides the motivations for creating the concentration
risk indictor (CRI) and how it is specially suited for the blockchain
landscape. Section (\ref{sec:A-Concentrated-Description}) gives a
high level overview of the concentration risk indicator including
the story of how - and more importantly why - the CRI was created.
Section (\ref{sec:Fund-Performance}) gives detailed mathematical
formulations for the CRI and other fund metrics that can be used along
with the CRI. Section (\ref{subsec:Multi-Chain-Concentration-Risk})
has an extension of the CRI when there are investments across multiple
chains. 

Section (\ref{sec:Numerical-Results}) explains the numerical results
we have obtained, which illustrate how our innovations compare to
existing wealth management techniques. Section (\ref{sec:Risk-Mitigation-Guidelines})
has a discussion of the challenges in risk management and some suggestions
on how to overcome them from an organizational point of view. Section
(\ref{sec:Risk-Mitigation-Guidelines}) acts as concise guide to financial
risk management by listing the five foundational pillars of risk management.
Section (\ref{sec:Pole-Vaulting-Over}) briefly describes the need
for continuous assessment and improvements in portfolio selection,
risk management and the entire circle of investment. Sections (\ref{sec:Areas-for-Further};
\ref{sec:Conclusion}) suggest further avenues for improvement - including
applications towards insurance and product portfolio mix risk mitigation
- and the conclusions respectively. 

\section{\label{sec:A-Concentrated-Description}Intuitive Insights: A Concentrated
Description of The Concentration Risk Indicator}

In this section, we provide intuitive insights regarding this innovation
and how the CRI metric was created to fill a gap in assessing portfolio
risks pertaining to the concentration of holdings. The development
of this metric provides a transparent way for clients to understand
how well a portfolio is being administered - in terms of risk and
diversification - which should send a powerful and lucid message to
- prospective and present - customers.

\subsection{\label{subsec:The-Origin-Story:}The Origin Story: Etiological Lore
of The CRI and Mixing Marketing with Financing}

At this juncture, it is important to describe how the CRI came about.
Marketing professionals - artificial as these distinctions between
disciplines are; this is a topic that can make for a long discussion
but is covered briefly in Kashyap (2021); End-note (\ref{enu:Artificial-Disciplines})
- in investment funds are faced with a unique problem. Investor relations
personnel are looking for some way to convey to investors - both existing
and potential ones - that diversified portfolios - and investment
strategies - are a much better risk adjusted means to earn high returns
instead of just holding cryptocurrencies such as Bitcoin or Ethereum
or one of the major tokens, either by themselves or a few of them
held together as a portfolio. 

Simply measuring the returns and risk - volatility - of an investment
- while helpful - do not convey the nuances of any concentrated portfolio
- or of investing in the blockchain realm. Bitcoin has a pretty decent
risk adjusted return compared to other securities in the Blockchain
space and would seem quite sufficient on its own - or held with a
few other tokens - as an investment. And most blockchain investors
would rather just hold Bitcoin or a few other crypto securities -
to get a decent risk return profile - and not take on the additional
risk of holding several smaller securities and the hassles of managing
many of them. The combinations of several securities can yield identical
risk and return profiles, though the full extent of risks are not
exactly similar due to concentration issues. The key is the tradeoff
when adding other securities to any portfolio and knowing which ones
are better to combine in a practical manner. 

The need is to articulate this concentration problem to existing blockchain
investors who hold large positions in Bitcoin and other tokens. A
new metric had to be developed that would convey the benefits of diversification
in a portfolio - or the lack of concentration - while adjusting for
the fact that large market capitalization securities do have a certain
amount of stability and are preferred over relatively smaller market
capitalization securities. This extra information is needed in addition
to risk and return accounting that has to be taken care of. The risk
of securities do capture the impact of their size, but in a portfolio
- where securities are combined - this message is diluted or even
entirely lost. While the effect of size has to be present, the metric
still had to weave in the risk and return profiles of investments,
otherwise we would be walking away from years of experience and wisdom.
The goal was to combine the risk and returns of securities with relative
size and portfolio weights. Thus, the CRI happened.

Hence, this origin story clarifies the very important point that this
paper contributes significantly to the finance and risk management
areas in addition to the marketing analytics literature. The creation
of the CRI is about going beyond the use analytics to check the effectiveness
of any marketing campaign and is about creating metrics that can help
to reveal the benefits of any product or service in a straightforward
yet prominent way. In this case the simplicity of the metric illustrates
how a powerful succinct communique can be crafted even for technological
products with many complex technical aspects. The main summary of
this metric is that the most effective and striking way to do marketing
is merely proper education with the right educational material. 

\subsection{\label{subsec:The-CRI:-A-Descendant}The CRI: A Descendant of The
HHI}

The concentration risk indicator is meant to indicate how diversified
the holdings in a portfolio are. This is a modification of the Herfindahl–Hirschman
(HH) index, (Rhoades 1993; Laine 1995; Sun \& Baker 2015; Lu, Qiao
\& Chang 2017; Sharma, Tiwari \& Rao 2020; End-note \ref{enu:The-Herfindahl-index}),
which is widely used as a measure of the size of firms in relation
to the industry they are in and an indicator of the amount of competition
among them. The HH index, like everything else, can be used or abused
(Van Houwelingen 1988; Nietzsche 2019). Despite its simplicity, or
perhaps even due to it, there are many instances where the HH index
can be improperly applied. Hence a proper understanding of the the
HH index, and specific scenarios where it is being applied, is necessary
before using it (Matsumoto, Merlone \& Szidarovszky 2012; Djolov 2013;
Brezina et al., 2016; Kvålseth 2018). 

There are many cases of the HH index being applied outside business,
sometimes with modifications, such as: to measure the competitive
balance in professional team sports, the degree of concentration of
container port systems, to derive the relative weights of multiple
non-commensurable transportation performance criteria, to evaluate
energy security risks, to examine the degree of repetition of individuals’
choices of their daily activity–travel–location combinations and even
the diversity of crops in agriculture (Owen, Ryan \& Weatherston 2007;
Le Coq \& Paltseva 2009; Le \& Ieda 2010; Susilo \& Axhausen 2014;
Bharati, De \& Pal 2015; Truong et al., 2021; Merk \& Teodoro 2022).

We tailor the HH-Index to the crypto markets based on the following
two features: 1) the larger the market cap of an asset, the lesser
the risk of holding it; 2) the more volatile an asset, the higher
the risk of holding it. The amount of money invested in an asset as
a fraction of the overall wealth held by an investor, which is also
the weight of the asset within the portfolio, is also factored in
this metric. The concentration risk indicator can be calculated for
individual assets and for portfolios of assets as well. When comparing
two investments, the lower the concentration risk, the better the
investment from a diversification point of view. If two assets have
comparable market cap then the asset with lower price volatility would
be preferred. Instead of using the raw market cap values, we normalize
and express it as a fraction to the total crypto market cap before
including this factor in the concentration risk indicator. Similarly,
if two assets have comparable levels of volatility, the asset with
a greater share of the market would be preferred.

As an illustration - given a choice between holding BTC or ETH - if
we need to isolate the effect of size on our investment, BTC with
its higher market capitalization would seem like a better alternative
(End-note \ref{enu:Crypto-Ranking}). Likewise, SOL, XRP and ADA have
a similar level of market share and hence their price volatility determines
how concentrated an investment in these assets would be. This simplified
example is meant only to illustrate the influence of size and volatility.
Clearly, ETH has many other features that could potentially qualify
it as a more desirable investment than BTC. An argument can also be
made that tokens with higher market cap will have lower volatility
than the ones with lower market cap (Fama \& French 1992; Perez‐Quiros
\& Timmermann 2000; Van Dijk 2011; Fama \& French 2018).

The first draft of this article originally included LUNA along with
SOL, XRP and ADA in the above example. The recent LUNA / UST episode
on the Terra network, from May 8 to May 13 2022 and beyond, is a demonstration
of the risk of holding concentrated portfolios (Uhlig 2022; Lee et
al. , 2022; Briola et al. , 2023).The events of May 2022 are a wake
up call to all players in the Crypto landscape. Better risk management,
stress testing and checking numerous seemingly unlikely scenarios
are an absolute necessity. High volatility episodes that occur during
financial crisis and corporate governance failures can have devastating
effects on portfolios with concentrated holdings (Ang \& Bekaert 2004;
Veldkamp 2005; Acharya \& Richardson 2009; Bates 2012; Fassin \& Gosselin
2011; Asad et al., 2020; Sheikh et al., 2020; Cole et al., 2021).

We have been developing and testing this new CRI metric for several
months before the collapse of LUNA. The creation of this measure was
to have a numeric score to show investors that no concentrated holding
is safe even if the position is on an asset as large as BTC or ETH.
Clearly the recent events, surrounding LUNA, have not been easy for
many of us. But it affirms our long held belief that nothing can be
taken for granted in crypto, and for that matter anywhere, and suitable
risk mitigation plans have to be made even for rather extreme scenarios.
These beliefs are encoded in the risk management guidelines, espoused
in Kashyap (2022), that investment teams have to adhere to. Going
beyond just this new metric, a rigorous approach to investing and
risk management is what investing on blockchain needs. Risk parity
and the whole suite of tools we are describing are exactly the need
of the hour.

\section{\label{sec:Fund-Performance}Methodologies for Measuring Fund Metrics:
Return, Volatility and Concentration Risk}

One approach to create Risk Parity would be to engineer a set of four
indexes or funds: Alpha, Beta, Gamma and Parity (ABGP: Kashyap 2022).
Alpha, Beta and Gamma are funds with different levels of risk and
expected returns. The investment mandates for these three funds will
be to ensure that, under most circumstances, Alpha will be more risky
than Beta and Beta will be more risky than Gamma. Investors will be
able to combine the three funds depending on their risk appetites.
Alpha, Beta and Gamma will be comprised of other crypto assets native
to the blockchain. environment Mixing Alpha, Beta and Gamma will give
the Risk Parity portfolio. 

The metrics show in Figure (\ref{fig:Performance-Reporting:-Risk,};
\ref{fig:Annualized-Risk-and}) would be good indicators for any investment
asset including for a portfolio of assets. That is we report return,
volatility and the concentration risk indicator for each of the three
sub funds Alpha, Beta and Gamma, used to construct Parity, and for
several individual crypto tokens or platforms. For investors we will
also show the realized and unrealized Profit and Losses (P\&L). Parity
will need additional considerations, which are covered in Section
(\ref{subsec:Fund-of-Fund}) below, and will be revisited separately
as well in later papers.

\subsection{\label{subsec:The-Fundamental-Figures:}The Fundamental Figures:
Risk and Return}
\begin{enumerate}
\item Asset Return - The formula for asset returns is given in Section (\ref{subsec:Return-and-Risk})
and Point (\ref{enu:Continously-Compounded-Return,}). For the 30
Day Return, $R_{t,30}$, the formula will become, 
\begin{equation}
R_{t,30}=\ln\left(\frac{P_{t}}{P_{t-30}}\right)
\end{equation}
\item Asset Volatility - The formula for asset volatility is given in Section
(\ref{subsec:Return-and-Risk}) and Point (\ref{enu:Volatility-at-time}).
For the 30 Day volatility, $\sigma_{t,30}$, the formula will become,
\begin{equation}
\sigma_{t,30}=\sigma_{t}\sqrt{30}
\end{equation}
Here, $\sigma_{t}$ is the daily volatility calculated in Section
(\ref{subsec:Return-and-Risk}) and Point (\ref{enu:Volatility-at-time}).
\item Realized and Unrealized Profits and Losses (P\&L) - We denote by $A_{t}$the
sequence of trade sizes or the amount of the asset traded at the corresponding
sequence of prices, $P_{t}$. Note that $A_{t}>0$ indicates a buy
trade and $A_{t}<0$ indicates a sell trade. Note that we are not
considering short sells here, which would require further modifications
to the formulae derived below. The total position, $POS_{T}$, at
a time $T$ is given by the sum of all the trades made until that
time.
\begin{equation}
POS_{T}=\sum_{t=1}^{T}A_{t}
\end{equation}
Here, $T$ is the total duration of trading under consideration. It
can also be the current time starting from an earlier time period.
For simplicity, time is measured in unit intervals giving, $t=1,2,...,T$.
We can also think of this as a sequence of trades indexed by $t=1,2,...,T$.
The weighted average buy price is given by, 
\begin{equation}
WAVGP_{T}=\frac{\sum_{t=1}^{T}\left\{ \mathbf{1}(A_{t}>0)\right\} A_{t}P_{t}}{\sum_{t=1}^{T}\left\{ \mathbf{1}(A_{t}>0)\right\} A_{t}}
\end{equation}
\begin{equation}
A_{0}=P_{0}=WAVGP_{0}=0
\end{equation}
$\left\{ \mathbf{1}(A)\right\} $ is the indicator function, which
gives $1$ if condition $A$ is TRUE or $0$ otherwise. 
\begin{equation}
{\displaystyle \mathbf{1}(A):={\begin{cases}
1~ & {\text{ if }}~A\text{ is \textbf{TRUE}}~,\\
0~ & {\text{ if }}~A\text{ is \textbf{FALSE}}~.
\end{cases}}}
\end{equation}
The implementation of weighted average uses the last time period weighted
average price with the corresponding quantity balance and the current
period quantity purchased with the current price. Such an approach
can be a more practical way to incorporate this in any system so that
the overall amount of calculations are reduced. The weighted average
price when calculated on a rolling basis will get adjusted for redemptions
(sell trades or withdraws) as follows,
\begin{equation}
WAVGP_{T}={\begin{cases}
\left(\frac{WAVGP_{T-1}POS_{T-1}+A_{T}P_{T}}{POS_{T-1}+A_{T}}\right)~ & {\text{ if }}~A_{T}>0~,\\
WAVGP_{T-1}~ & {\text{ if }}~A_{T}<0~.
\end{cases}}
\end{equation}
The Unrealized P\&L at a certain time $T$ is given by,
\begin{equation}
PANDLUNREAL_{T}=\left(POS_{T}\right)\left(P_{T}-WAVGP_{T}\right)
\end{equation}
The Realized P\&L until a certain time $T$ is given by,
\begin{equation}
PANDLREAL_{T}=\sum_{t=1}^{T}\left\{ \mathbf{1}(A_{t}<0)\right\} \left|A_{t}\right|\left(P_{t}-WAVGP_{t}\right)
\end{equation}
The realized P\&L can only start from the second time period or from
the second set of trades since the first trade has to be a buy trade
since we are not considered short sales here. That is, the first time
period realized P\&L is zero. 
\begin{equation}
PANDLREAL_{1}=0
\end{equation}
For a practical implementation we can use the following rolling formulae
for the Realized P\&L,
\begin{equation}
PANDLREAL_{T}={\begin{cases}
PANDLREAL_{T-1}~ & {\text{ if }}~A_{T}>0~,\\
PANDLREAL_{T-1}+\left|A_{T}\right|\left(P_{T}-WAVGP_{T}\right)~ & {\text{ if }}~A_{T}<0~.
\end{cases}}
\end{equation}
The Unrealized Return at a certain time $T$ is given by,
\begin{equation}
RETURNUNREAL_{T}={\begin{cases}
\left(\frac{PANDLUNREAL_{T}}{WAVGP_{T}POS_{T}}\right)~ & {\text{ if }}~POS_{T}>0~,\\
0~ & {\text{ if }}~POS_{T}=0~.
\end{cases}}
\end{equation}
The Realized Return until a certain time $T$ is given by,
\begin{equation}
RETURNREAL_{T}={\begin{cases}
RETURNREAL_{T-1}~ & {\text{ if }}~A_{T}>0~,\\
\left(\frac{PANDLREAL_{T}}{WAVGP_{T-1}POS_{T-1}}\right)~ & {\text{ if }}~A_{T}<0~.
\end{cases}}
\end{equation}
The realized return can only start from the second time period or
from the second set of trades since the first trade has to be a buy
trade since we are not considered short sales here. That is, the first
time period realized return is zero.
\begin{equation}
RETURNREAL_{1}=0
\end{equation}
We can also have separate variables for buy and sell trades to obtain
results equivalent to the above.
\end{enumerate}

\subsection{\label{subsec:Asset-Concentration-Risk}Asset or Portfolio Concentration
Risk Indicator}
\begin{enumerate}
\item The concentration risk is meant to indicate how diversified the holdings
in a portfolio are. This is a modification of the Herfindahl–Hirschman
(HH) index: (\href{https://en.wikipedia.org/wiki/Herfindahl\%E2\%80\%93Hirschman_index}{Herfindahl–Hirschman index}),
which is widely used as a measure of the size of firms in relation
to the industry they are in and an indicator of the amount of competition
among them. We tailor the HH Index to the crypto markets based on
the following two features: 1) the larger the market cap of an asset,
the lesser the risk of holding it. 2) the more volatile an asset,
the higher the risk of holding it. The concentration risk indicator
can be calculated for individual assets and also for portfolios of
assets. The formula is given below,
\begin{equation}
CRI_{t}=\left(\frac{1}{k_{t}}\right)\sum_{i=1}^{k_{t}}\left(\frac{\sigma_{it}}{m_{it}}\right)\left(w_{it}\right)^{2}\label{eq:CRI-Score}
\end{equation}
Here, $CRI_{t}$ is the concentration risk indicator for a portfolio
with $k_{t}$ assets at time $t$. $\sigma_{it}$ is the volatility
of asset $i$ at time $t$. $w_{it}$ is the weight of asset $i$
in the overall portfolio at time $t$. $m_{it}$ is the ratio of the
market capitalization of asset $i$ to the total market capitalization
of the entire market at time $t$. That is,
\begin{equation}
m_{it}=\left(\frac{MC_{it}}{TMC_{t}}\right)\label{eq:Market-Factor}
\end{equation}
Here, $MC_{it}$ is the market capitalization of asset $i$ at time
$t$. $TMC_{t}$ is the total market capitalization of the entire
market at time $t$. For a portfolio with only one asset such as BTC
or ETH, note that $k_{t}=w_{it}=1$.
\end{enumerate}

\subsection{\label{subsec:Fund-of-Fund}Fund of Fund Extension: Parity Risk,
Return and CRI}

The results for Parity is the result of combining the metrics for
Alpha, Beta and Gamma. Hence, we can consider this as a fund of fund
extension.
\begin{enumerate}
\item \label{enu:Parity-Return,-Volatility}Parity Return, Volatility and
Risk Concentration Indicator - Kashyap (2022) has further details
on Parity portfolio construction. Let the weights of Alpha, Beta and
Gamma funds in a Parity portfolio be $\left(w_{\alpha},w_{\beta},w_{\gamma}\right)$.
The weights could be based on the choice of any investor or it could
be for the three characteristic funds: Parity Low, Parity Medium and
Parity High which specify default values for low, medium and high
risk preferences. Note that the weights satisfy the equation, 
\begin{equation}
w_{\alpha}+w_{\beta}+w_{\gamma}=1
\end{equation}

\begin{enumerate}
\item The return for Parity will be given by the following equation,
\begin{equation}
R_{P}=R_{\alpha}w_{\alpha}+R_{\beta}w_{\beta}+R_{\gamma}w_{\gamma}
\end{equation}
Here, $\left(R_{P},R_{\alpha},R_{\beta},R_{\gamma}\right)$ are the
returns for Parity, Alpha, Beta and Gamma respectively. 
\item The volatility for Parity will be given by the following equations,
\begin{equation}
\sigma_{P}=\sqrt{\sigma_{\alpha}^{2}w_{\alpha}^{2}+\sigma_{\beta}^{2}w_{\beta}^{2}+\sigma_{\gamma}^{2}w_{\gamma}^{2}+2w_{\alpha}w_{\beta}\sigma_{\alpha}\sigma_{\beta}\rho_{\alpha,\beta}++2w_{\alpha}w_{\gamma}\sigma_{\alpha}\sigma_{\gamma}\rho_{\alpha,\gamma}++2w_{\gamma}w_{\beta}\sigma_{\gamma}\sigma_{\beta}\rho_{\gamma,\beta}}
\end{equation}
An equivalent result using the covariance matrix is given below,
\begin{equation}
\boldsymbol{\sigma_{p}}=\boldsymbol{w'Xw}
\end{equation}
 Here, $\left(\sigma_{P},\sigma_{\alpha},\sigma_{\beta},\sigma_{\gamma}\right)$
are the volatilities for Parity, Alpha, Beta and Gamma respectively.
$\left(\rho_{\alpha,\beta},\rho_{\alpha,\gamma},\rho_{\gamma,\beta}\right)$
are the correlations between Alpha and Beta, Alpha and Gamma, and
Beta and Gamma respectively. $\boldsymbol{w}$ is the weight row vector
with weights for Alpha, Beta and Gamma. $\boldsymbol{w'}$ is the
weight column vector with weights for Alpha, Beta and Gamma. $\boldsymbol{X}$
is the covariance matrix for Alpha, beta and Gamma. 
\item The concentration risk indicator for Parity will be built bottom up
using the original CRI formulation in Equation (\ref{eq:CRI-Score}).
The weights of the assets in Alpha, Beta and Gamma for the Parity
CRI formulation are changed by multiplying them with the corresponding
weights of Alpha, Beta and Gamma in Parity. For example, the weights
of assets in Alpha for the CRI score will be modified as follows,
\begin{equation}
wmod_{it}=\left(w_{it}\right)\left(w_{\alpha}\right)
\end{equation}
Here, $\left(wmod_{it}\right)$ is the modified weight of asset $i$
in Alpha for the Parity CRI score. A similar adjustment is done for
all the assets in Alpha, Beta and Gamma. Let $K_{t}$be the sum of
the total number of assets in Alpha Alpha, Beta and Gamma. We then
substitute the modified individual asset weights and the total number
of assets accordingly in the original CRI Equation (\ref{eq:CRI-Score})
to get the Parity CRI as below,
\begin{equation}
CRI_{t}=\left(\frac{1}{K_{t}}\right)\sum_{i=1}^{K_{t}}\left(\frac{\sigma_{it}}{m_{it}}\right)\left(wmod_{it}\right)^{2}
\end{equation}
\end{enumerate}
\end{enumerate}

\subsection{\label{subsec:Multi-Chain-Concentration-Risk}Multi-Chain Concentration
Risk Indicator }

When we have multiple chains, $N_{C}=$ number of chains, it might
be good to derive a score that considers the proportion of assets
deployed on each chain. Hence we need to add additional terms, that
factor the number of chains in the above formulation (Eq: \ref{eq:CRI-Score}).
This enhancement will be more thoroughly done in a later iteration.
But the following multiplicative factors can be useful next steps.
In the first formulation below, if each network carries the same proportion
of the total wealth, then the multiplicative factor is one. Any difference
in the proportions from equality of each leads to a higher multiplicative
factor and will increase the concentration risk indicator in Equation
( \ref{eq:CRI-Score}).
\begin{equation}
MCF_{t}=\left[\left(\frac{1}{N_{C}}\right)\left\{ \sum_{i=1}^{N_{C}}\left(\frac{\left(\frac{1}{N_{C}}\right)}{CW_{it}}\right)\right\} \right]\quad;\qquad\sum_{i=1}^{N_{C}}CW_{it}=1
\end{equation}
This can be simplified to,
\begin{equation}
MCF_{t}=\left[\left\{ \frac{1}{\left(N_{C}\right)^{2}}\right\} \left\{ \sum_{i=1}^{N_{C}}\left(\frac{1}{CW_{it}}\right)\right\} \right]\label{eq:Multi-Chain-Distribution}
\end{equation}
Here, $MCF_{t}$ is the multiple chain factor at time $t$. $CW_{it}$
is the proportion of assets deployed on chain $i$ at time $t$. Instead
of using the ratio, $\left(\frac{1}{N_{C}}\right)$ to indicate that
the assets are evenly distributed among the different networks, we
can use the proportion of the market capitalization of any network
as compared to the total market capitalization across the networks
in considerations. $MC_{it}$ is the market capitalization of network
$i$ at time $t$. 
\begin{equation}
MCF_{t}=\left[\left(\frac{1}{N_{C}}\right)\left\{ \sum_{i=1}^{N_{C}}\left(\frac{\left(\frac{MC_{it}}{\sum_{i=1}^{N_{C}}MC_{it}}\right)}{CW_{it}}\right)\right\} \right]
\end{equation}
This can be simplified to,
\begin{equation}
MCF_{t}=\left[\left(\frac{1}{N_{C}}\right)\left(\frac{1}{\sum_{i=1}^{N_{C}}MC_{it}}\right)\left\{ \sum_{i=1}^{N_{C}}\left(\frac{MC_{it}}{CW_{it}}\right)\right\} \right]\label{eq:Multi-Chain-Distribution-MarketCap}
\end{equation}
The benefit of additional chains can be captured by multiplying Equations
(\ref{eq:Multi-Chain-Distribution}; \ref{eq:Multi-Chain-Distribution-MarketCap})
with an additional $\left(\frac{1}{N_{C}}\right)$ factor. This ensures
that adding more chains reduces the concentration risk indicator further
and is representative of the greater diversity in the holdings.

\subsection{\label{subsec:Return-and-Risk}Return and Risk (Volatility) Calculations}

The return will be the continuously compounded measure. We start with
daily returns, though we could use other frequencies as well. The
volatility will be the standard deviation of the continuously compounded
returns over a historical time period, generally around 90 days. The
volatility will need to be calculated on a rolling 90 days basis. 
\begin{enumerate}
\item \label{enu:Continously-Compounded-Return,}Continuously Compounded
Return, $R_{t}$, at any time $t$ is given by the logarithm of the
ratio of the Price at time $t$, $P_{t}$, and the Price at time $t-1$,
$P_{t-1}$, as shown below,
\begin{equation}
R_{t}=\ln\left(\frac{P_{t}}{P_{t-1}}\right)
\end{equation}
\item \label{enu:Volatility-at-time}Volatility at time $t$, $\sigma_{t}$
is calculated as below using the average return, $\bar{R}_{t}$, at
time $t$ and then calculating the standard deviation. It can be done
directly if suitable libraries are available. $\sigma_{t}^{2}$ is
the variance at time $t$. Here, $T=90$ unless otherwise stated.
\begin{equation}
\bar{R}_{t}=\left(\frac{1}{T}\right)\sum_{i=t}^{t-T}R_{i}\label{eq:Return-Estimate}
\end{equation}
\begin{equation}
\sigma_{t}^{2}=\left(\frac{1}{T-1}\right)\sum_{i=t}^{t-T}\left(R_{i}-\bar{R}_{t}\right)^{2}
\end{equation}
\begin{equation}
\sigma_{t}=\sqrt{\left(\frac{1}{T-1}\right)\sum_{i=t}^{t-T}\left(R_{i}-\bar{R}_{t}\right)^{2}}\label{eq:Volatility-Risk-Estimate}
\end{equation}
\item Calculate the volatility for the longest historical time period possible
for each asset. Possibly for the last 360 days or longer. Some new
assets might have a relatively shorter price time series, and it is
okay to calculate volatility for those assets for the number of days
for which prices are available.
\end{enumerate}

\section{\label{sec:Numerical-Results}Data Analysis and Numerical Results}

Each of the tables in this section are referenced in the main body
of the article. Below, we provide supplementary descriptions for each
table. 
\begin{itemize}
\item The Table in Figure (\ref{fig:Performance-Reporting:-Risk,}) shows
numerical examples related to the metrics described in Section (\ref{sec:Fund-Performance}). 
\item The dataset we have used is publicly available from Crypto market
data providers such as CoinMarketCap (End-note \ref{enu:CoinMarketCap}).
The data was collected into a SQL Relational Database using a suitable
Application Programming Interface (API) provided by CoinMarketCap
(Drobetz et al., 2019; Harrington 2016; Perez et al., 2020; Parlika
\& Pratama 2021; Jacobson et al., 2012; Meng et al., 2018; End-note
\ref{enu:Structured-Query-Language}; \ref{enu:An-application-programming}).
Some of the calculations were done on the database using the corresponding
database query language and some were done external to the database
using python utilities (Van Rossum \& Drake 1995; End-note \ref{enu:Python-is-a}).
\item The metrics in Figure (\ref{fig:Performance-Reporting:-Risk,}) are
estimates calculated based on daily values from January 01, 2019 until
August 10, 2022.
\item Alpha consists of about 24 securities with high market capitalization
(End-note \ref{enu:Market-capitalization,-sometimes}). Beta consists
of 60 securities, and is more representative of the wider market than
Alpha, including several assets with lower market capitalization.
Beta can be grouped into several thematic sectors such as NFTs, Oracles,
Networks, DeFi, etc. Gamma consists of low volatility instruments
such as stable coins (Ante, Fiedler \& Strehle 2021; End-note \ref{enu:Stablecoin}).
We see from Figures (\ref{fig:Performance-Reporting:-Risk,}; \ref{fig:Annualized-Risk-and})
that a combined portfolio of crypto assets can beat BTC in terms risk,
return and CRI. This confirms the intuition that diversification and
better risk management are absolutely necessary for crypto investment
management.
\item The columns in Figure (\ref{fig:Performance-Reporting:-Risk,}) represent
the following information respectively: 
\begin{enumerate}
\item \textbf{AssetName} is the name of the asset or portfolio for which
the metrics, in the next three columns, are reported in this row.
\item \textbf{30 Day Return }is the estimate of the 30 day expected return,
calculated using an arithmetic average of returns over the data sample.
\item \textbf{30 Day Volatility} is the estimate of the 30 day historical
volatility calculated using daily continuously compounded returns
over the data sample.
\item \textbf{Concentration Risk} is the CRI using the 30 day historical
volatility. 
\end{enumerate}
\end{itemize}
\begin{doublespace}
\begin{figure}[H]
\includegraphics[width=18cm]{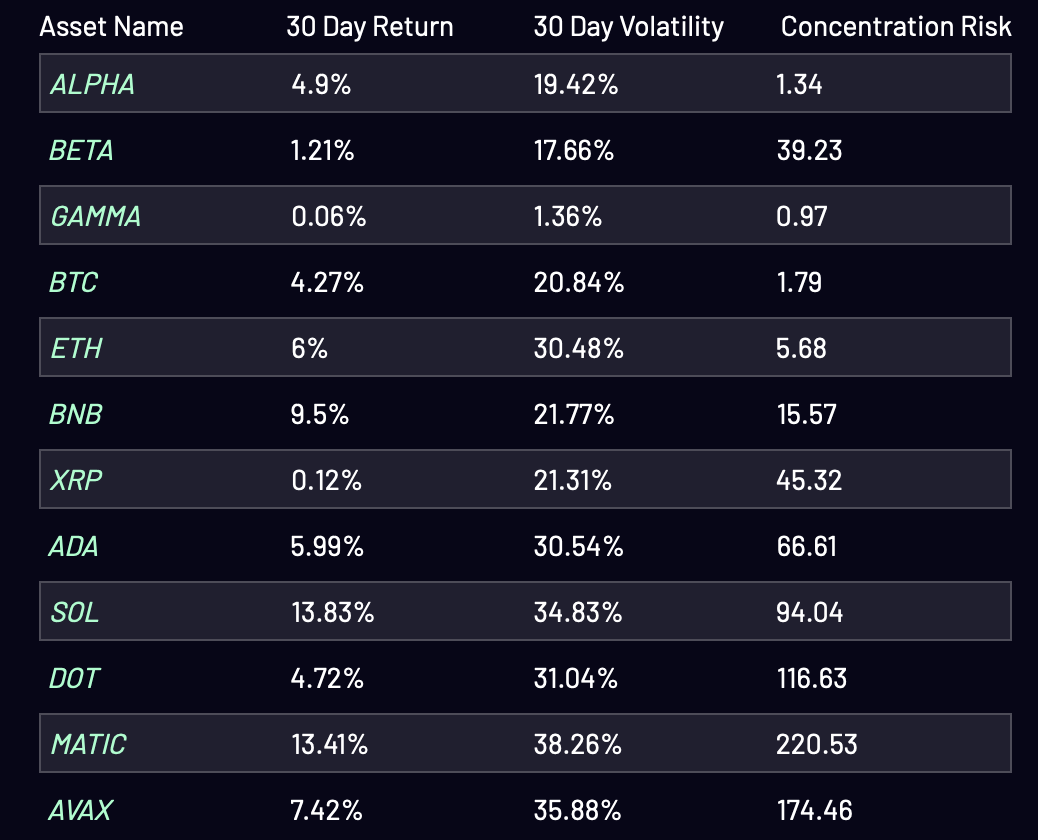}

\caption{\label{fig:Performance-Reporting:-Risk,}Performance Reporting: Return,
Volatility and Concentration Risk Indicator}
\end{figure}

\end{doublespace}
\begin{itemize}
\item The Table in Figure (\ref{fig:Annualized-Risk-and}) shows numerical
examples related to the method described in Section (\ref{sec:Fund-Performance}). 
\item The dataset we have used is publicly available from Crypto market
data providers such as CoinMarketCap (End-note \ref{enu:CoinMarketCap}).
\item The metrics in Figure (\ref{fig:Annualized-Risk-and}) are estimates
calculated based on daily values from January 01, 2019 until Aug 10,
2022. 
\item The columns in Figure (\ref{fig:Annualized-Risk-and}) represent the
following information respectively: 
\begin{enumerate}
\item \textbf{AssetName} is the name of the asset or portfolio for which
the metrics, in the next three columns, are reported in this row.
\item \textbf{30 Day Return }is the estimate of the annual expected return,
calculated using an arithmetic average of returns over the data sample.
\item \textbf{30 Day Volatility} is the estimate of the annual historical
volatility calculated using daily continuously compounded returns
over the data sample.
\item \textbf{Concentration Risk} is the CRI using the 30 day historical
volatility. 
\item \textbf{Risk Adjusted Return} is calculated as the ratio of the difference
between Asset Return and Risk Free Rate divided by the Asset Volatility.
The Risk Adjusted Return is calculated based on a Risk Free Rate assumption
of 8\%. 
\end{enumerate}
\end{itemize}
\begin{doublespace}
\begin{center}
\begin{figure}[H]
\includegraphics[width=18cm]{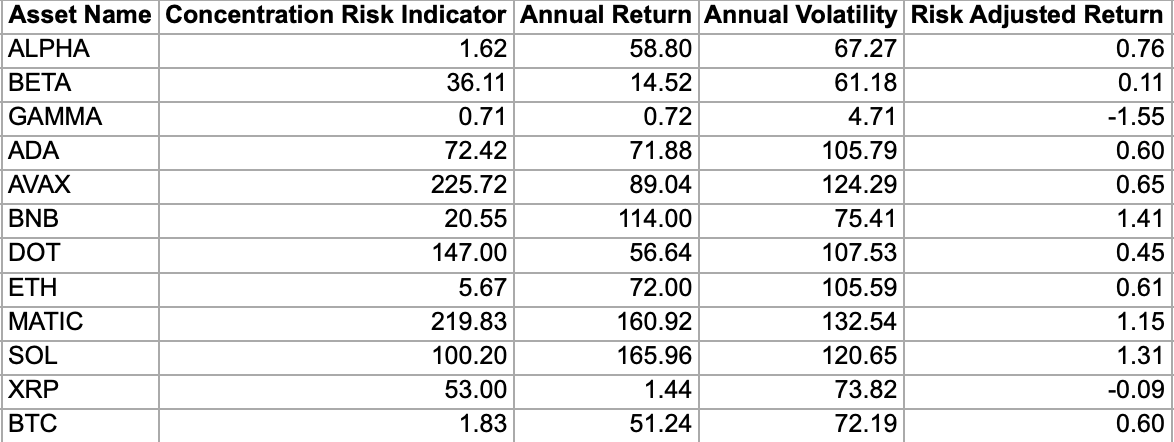}

\caption{\label{fig:Annualized-Risk-and}Early Results of CRI, Annualized Return,
Annualized Volatility and Risk Adjusted Return}
\end{figure}
\par\end{center}
\end{doublespace}
\begin{doublespace}

\section{\label{subsec:A-Slow-Walk}Discussion: Risk Management, A Slow Walk
On A Tight Rope …}
\end{doublespace}

\begin{figure}[H]
\includegraphics[width=18cm]{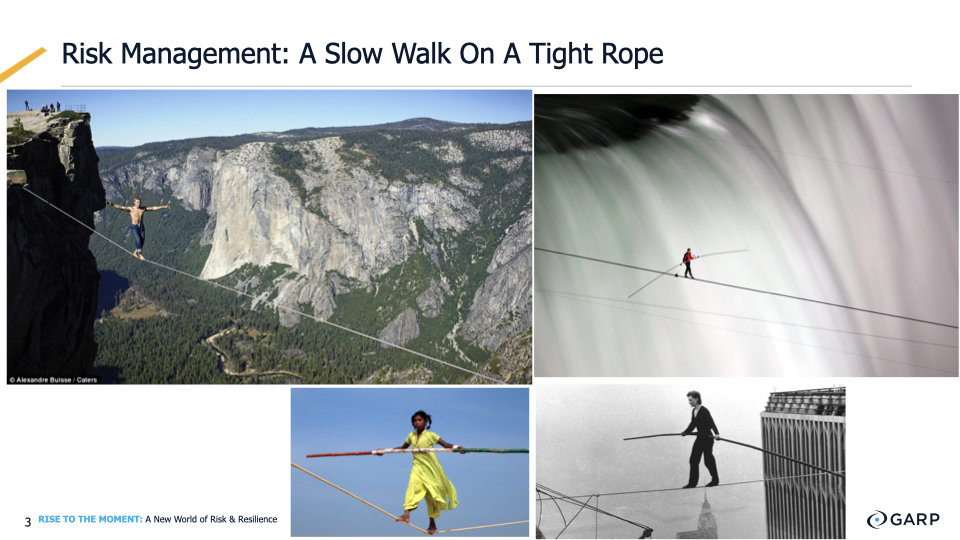}\caption{\label{fig:Risk-Management-Slow-Walk-Tight-Rope}Risk Management:
A Slow Walk On A Tight Rope …}
\end{figure}

\begin{doublespace}
All the metrics we create, the models we develop and the measurements
we make using actual data could be pointless unless the foundations
of risk management are on a strong footing. Hence, as an aside, it
is worth looking more closely at the role of a Risk Manager. This
discussion also applies to some extent - albeit lesser extent - to
compliance officers, controllers and auditors (Gupta 1987; Jorion
2007; Horcher 2011; Graham, Davey‐Evans \& Toon 2012; Lin 2016). 

Financial trading firms have elaborate structures of people assigned
to monitor and support the activities of trading desks. Most of the
support functions are handled by operations teams in terms of booking
trades, ensuring trades are cleared on the exchanges or settlement
houses and the reporting into the technology infrastructure is done
correctly and on time. The monitoring roles can be broadly viewed
as: controllers, compliance officers and risk managers, while keeping
in mind that there are significant overlaps in the day to day tasks
and responsibilities of these groups of individuals (Mikes 2008; Lim
et al., 2017). 

Risk managers focus on obtaining numeric measures indicating the extent
of risk, or the potential for loss, due to trading activities (Alexander
2005; Christoffersen 2011; Malz 2011; End-note \ref{enu:Financial-risk-management}).
Controllers focus on ensuring that trades are booked correctly, the
counter parties are valid organizations and that the Profits and Losses,
(P\&L) and related accounting is accurate and reported correctly (Hrisak
1996; Zoni \& Merchant 2007; End-note \ref{enu:A-financial-controller,}).
Compliance officers ensure compliance with current and expected regulatory
guidelines (Arjoon 2005; Griffith 2015; End-note \ref{enu:Regulatory-compliance-describes}).
Auditors can be both internet and external, and provide an independent
examination of the financial information of any organization (Beck
1973; Chandler, Edwards \& Anderson 1993; Brody \& Lowe 2000; End-note
\ref{enu:An-auditor-is}). Note that risk management outside of financial
risk is an extremely well studied and broad field with significant
bearings on all aspects of life (Covello \& Mumpower 1985; Colquitt,
Hoyt \& Lee 1999; Ward 2001; Power 2004; Hubbard 2020; End-note \ref{enu:Risk-management-is}).

If risk managers do their job well, it is hard to know the extent
of the risks they have averted. If there is a blow up - a huge loss
in a portfolio - of course, they have failed at their jobs. The one
- perhaps the only way? - they can do their jobs perfectly is by not
letting their traders / portfolio managers take on risky trades. But
then again risk managers get compensated by the very profits from
the P\&L of their traders that depends on taking on bigger risks.
So what risk managers really look for is some way to identify portions
of the portfolio they are risk managing using some criteria. The risk
managers then issue reports - using the criteria they have identified
- saying these are risky trades. Risk managers hope that if there
is a blow up at a later stage the risky trades are in those outlier
reports they had put out earlier. Let us just say their daily work
lives might feel like a slow walk on a tight rope (Landier, Sraer
\& Thesmar 2009; Figure \ref{fig:Risk-Management-Slow-Walk-Tight-Rope}).
An understanding of this primary riddle - related to risk management
that the risk manager is someone doing a slow walk on a tight rope
- will aid in creating the right set of governance mechanisms within
any organization.

Another aspect to consider are the motivations for someone to become
a trader - portfolio manager - or a risk manager. Any talented individual
starting a career in financial services would be more keen to become
a trader than a risk manager. This is because traders are generally
compensated more highly than risk managers on average - let us say
- for similar levels of experience and organizational responsibilities
(Koenig 2006; Chen \& Fabozzi 2010; Noe \& Young 2014; Van Boxtel
2017; Efing \& Kampkötter 2020). This leads to a bias within - and
even outside - the organization that someone that ends up as a trader
is more skilled. 

Viewing this from the flip side leads to the belief that someone ends
up as a risk manager because they could not become a trader. The implication
- possibly erroneous - is that whoever did not become traders were
bad at performing the role of traders or even because they lacked
the necessary skills to make complex decisions involved in deciding
which trades to execute. This perception is fueled regularly on trading
desks when traders routinely tell risk managers they do not understand
certain aspects of trading when asked about why a certain trade was
made. Sometimes this can even be via the use of condescending words
such as, ``you don't even know such simple things'' or ``how did
someone make you a risk manager?'' or ``I don't have the time to
do an Introduction - Introductory 101 Course - to finance for you
now''.

Nobody likes to be told that they cannot understand something and
certainly not that they are someone who is deemed to have less knowledge.
This bias means that many times risk managers do not challenge traders.
Risk managers do not try to pursue matters in depth when they fail
to comprehend certain things related to trading strategies they are
supposed to risk manage. We see a classic example of this happening
in Kashyap (2021b). 

We wish to emphasize that there are many excellent risk managers who
are absolutely passionate and extremely skilled at what they do. The
discussion here covers the norm - or the more common observations
- rather than the exception. To be fair to traders as well, they are
under a lot of stress working constantly in extreme pressure situations
to generate profits day in and day out. Making a big loss - or bad
investment decision - leads to immediate termination for traders.
Such stressful scenarios can lead to angry outbursts and flared tempers.
\end{doublespace}

\textbf{\textit{The way around this issue is to create a culture where
risk management is cherished. Surely, risk takers are to be respected
since taking risks is not natural to most of us. But an organization
that celebrates risk managers - who avert big crisis - along with
risk takers, that take bold steps into the unknown, is very much the
need of the hour.}}

\subsection{\label{sec:Risk-Mitigation-Guidelines}Five Pillars: Risk Mitigation
Guidelines}

It is tempting to create lengthy documents for risk management. While
detailed expositions are helpful what is necessary - as a first step
- is brevity so that it is readily apparent what actions to perform,
and implement, without much confusion. Comprehensive supplements can
be added as explanations for an essential set of ideals that need
to followed while managing assets - such as crypto assets or insurance
products. A core package of principles applies similarly to traditional
financial assets - or portfolios of different products and services
- as well. We list below a set of concise directives - the five pillars
risk mitigation guidelines - which are meant to provide clear pointers
on assessing financial portfolio risks. 

With the rapidly developing field of crypto risk management, these
guidelines would be highly suited for the greater volatility seen
in this landscape. A core package of principles, similar to what we
have outlined below, would readily apply to cryptocurrency assets
with minor alternations specific to the nuances of the crypto-portfolios
under consideration. 

The following guidelines are purposefully kept brief to facilitate
easy assimilation within any organization.
\begin{enumerate}
\item \textbf{\textit{\label{enu:The-recommended-total-weight}The recommended
total weight for any asset, across all funds, will be less than $X\%$. }}
\begin{itemize}
\item If any assets have to be allocated more than $X\%$, additional due
diligence has to be performed and further justifications need to be
considered. 
\item $Y\%$ is the absolute limit for any asset in terms of weight in the
portfolio. Note that, $Y\geq X$. 
\item The number of assets in the $X\%$ to $Y\%$ range should be less
than $N$. 
\item Suggested values for $X$, $Y$ and $N$ are 10\%, 15\% and 3 respectively.
Note that $N$ can be a certain percentile of the total number of
assets in the overall portfolio (End-note \ref{enu:Percentile}).
\end{itemize}
\item \label{enu:Unwind-Full}\textbf{\textit{As soon as we have lost confidence
in any asset, its position is to be closed out and the corresponding
amounts need to be gradually unwound. }}
\begin{itemize}
\item A quicker exit option can be pursued when there is a distress or panic
situation going on. This relates to Point (\ref{enu:The-total-daily})
which imposes a limit on how much we can sell (or buy) in a given
day depending on the daily trading volume (End-note \ref{enu:Trading-Volume}).
\end{itemize}
\item \textbf{\textit{If any asset price drops by more than $Z\%$, then
further buys will be done only on an exceptional basis with additional
due diligence and research to be performed. }}
\begin{itemize}
\item It is important to remember that when any asset price drops, and its
portfolio weight has not changed, then the rebalancing algorithm will
try to buy more of it (Kashyap 2022). 
\item The recommended approach when there is a $Z\%$, or bigger, drop is
to hold the asset. 
\item Furthermore, we need to consider if Point (\ref{enu:Unwind-Full})
becomes applicable and act accordingly. 
\item For stable coins that drop more than $S\%$, Point (\ref{enu:Unwind-Full})
becomes applicable on a high priority. 
\item Suggested values for $Z$ and $S$ are 25\% and 10\% respectively.
\end{itemize}
\item \textbf{\textit{The total position in any token, across all funds,
cannot be more than $M\%$ of the market capitalization of that token
(End-note \ref{enu:Market-capitalization,-sometimes}). }}
\begin{itemize}
\item Suggested values for $M$ is 5\%.
\item An exception can be granted to a small subset of securities in the
portfolio, similar to Point (\ref{enu:The-recommended-total-weight}),
in terms of their positions being greater than \textbf{\textit{$M\%$
}}of the market capitalization of that token. 
\end{itemize}
\item \label{enu:The-total-daily}\textbf{\textit{The total daily trading
volume in any token, across all funds, cannot be more than V\% of
the entire market trading volume of that token. }}
\begin{itemize}
\item The market trading volume is better considered as a daily average
over a suitable time period, such as a month. 
\item For exceptional investment circumstances, trading that involves higher
market volumes can be considered. These actions have to be based on
considerations that apply from Point (\ref{enu:Unwind-Full}). 
\item Suggested values for $V$ is 10\%.
\end{itemize}
\end{enumerate}
The suggested values we have given are based on historical variations
we have observed in the parameters across different asset classes
but particularly attuned to higher volatility instruments such as
crypto-currencies. Numerous quantitative methods can be used to arrive
at the estimates for the parameters. The parameters can be arrived
at, so that VaR, Conditional VaR and other measures can fall within
reasonable bounds, based on either historical simulations or monte-carlo
computation (Jorion 1996; 2006; Duffie \& Pan 1997; Linsmeier \& Pearson
2000; Pflug 2000; Rockafellar, R. T., \& Uryasev 2000; 2002; Trucíos,
Tiwari \& Alqahtani 2020; Peng, Yang \& Yao 2023; End-notes \ref{enu:Value-at-risk};
\ref{enu:Expected-shortfall-(ES)}).

\subsection{\label{sec:Pole-Vaulting-Over}Pole Vaulting Over Uncertainty}

\begin{figure}[H]
\includegraphics[width=18cm]{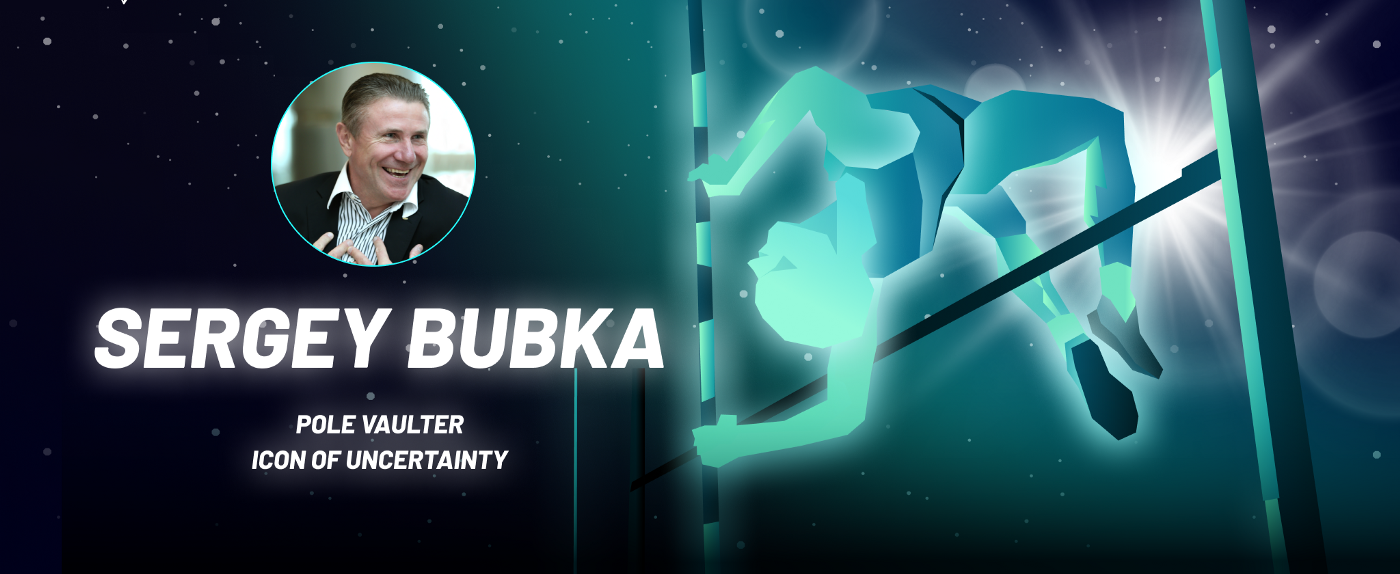}\caption{\label{fig:Sergey-Bubka:-Icon}Sergey Bubka: Icon of Uncertainty}
\end{figure}

A big part of investing, and also perhaps many other aspects of our
lives, is dealing with uncertainty and our struggle to overcome it.
Sergey Bubka is our Icon of Uncertainty (Figure \ref{fig:Sergey-Bubka:-Icon}).
As a refresher, he broke the pole vault world record 35 times (End-note
\ref{enu:Sergey-Nazarovich-Bubka}). Pole vault is a simple sport
wherein you use a long pole to jump over another long pole, which
is placed on top of two other long poles. 

Applying the central idea from pole vault to the investment and risk
mitigation landscape, we can view the introduction of any new trading
strategy - or risk model or metric or innovation or even regulatory
change - as equivalent to the raising of the bar in the game of pole
vault. Once a new innovation starts becoming popular others imitate
it, or come up with other wonderful ideas, and we need to find ways
to better ourselves. Each time the bar is raised the spirit of Sergey
Bubka, whom we admire a lot and who is a huge inspiration for us,
will help us to reach higher and find a way over the raised bar. 

This anecdote about Sergey Bubka, and overcoming uncertainty, forms
our fundamental belief that galvanizes us to constantly innovate and
find better models, metrics, trading strategies, risk mitigation techniques
and ways to generate wealth for all investment participants.

\section{\label{sec:Areas-for-Further}Areas for Further Research}

\subsection{\label{subsec:Insurance-Risk-and}Insurance Risk and Product Mix
Concentration}

A simple modification can make this metric very useful for the insurance
industry (Austin 1983; Cowley \& Cummins 2005; Cummins et al., 2013;
Asmussen \& Steffensen 2020; Shi \& Shi 2023; End-note \ref{enu:Insurance-is-a}).
We can consider different risk categories - such as different age
groups for life insurance or different localities for property protection
- as different assets in a portfolio. The weight of each risk category
would be the extent of liabilities in that category as compared to
the overall liabilities for the entire product mix. The volatility
of the asset would be the variation in the liability value under each
category over a period of time. 

The market factor we have used - in the CRI formulation given in Equation
(\ref{eq:Market-Factor}) - would be similar to comparing the total
liability held in a risk group in that particular portfolio compared
to the wider market. When using the market factor we need to take
the inverse ratio - as compared to the original formula in Section
(\ref{subsec:Asset-Concentration-Risk}) - since larger values in
the denominator are better. Similar adjustments can be useful for
gauging the risks for any product portfolio mix (Kahane 1977; Tsai
et al., 2010; McNally et al., 2013; Brzęczek 2020).

\subsection{\label{subsec:Potential-Improvements-to}Potential Improvements to
the CRI}

Several extensions to the CRI are possible. One immediately feasible
inquiry - we are engaged in - is to check which of the properties
of a coherent risk measure are satisfied by the CRI (Artzner et al.,
1999; Dhaene et al., 2008; Asimit \& Li 2016; End-note \ref{enu:Coherent-Risk-Measure}). 

We can consider the covariance between assets to come up with a new
indicator. It is important to realize that several mathematical results
can be developed to demonstrate how the CRI changes as different assets
are combined in a portfolio. Comparing the changes in the CRI to the
corresponding risk and return profiles would be a interesting exercise.
In particular, combining several assets can give identical risk and
return and hence checking the extent to which the CRI will give a
different value would be helpful.

While the CRI metric gives preference for larger and more stable assets,
smaller and newer assets will be the drivers of growth. Hence, adding
a greater number of smaller assets can compensate for the risk they
bring in terms of size. Asset selection guidelines, due diligence
processes, proper risk management oversight and the VVV weighting
methodology we describe in Kashyap (2022) take care of monitoring
the many other factors that dictate whether an asset makes a good
investment.As better blockchain networks develop, we will need to
see if the above techniques we have created need modification. 

We are deeply cognizant of the delicate necessity that to keep improving
the performance of our portfolios, our tools to manage risk and assess
performance need to keep improving as well. Volatility, returns and
other metrics are generally more meaningful when evaluated on a comparative
or relative basis. Since Crypto investments are deemed riskier - to
perform a proper comparison for risk and return - it will be helpful
to try to incorporate benchmarks external to the crypto world. Initially
it would be easier to start displaying returns, volatility and the
concentration risk indicator, over different time intervals. Such
comparisons across different time periods could check the metrics
across the main funds we have discussed - Alpha, Beta and Gamma -
with several other prominent crypto assets. Parity - which is the
combination of Alpha, Beta and Gamma funds - investments at different
risk levels can also be viewed as different crypto funds and similar
comparisons can be performed.

At a later stage, we can compare the volatility of crypto investment
funds to an external benchmark such as the VIX volatility index (End-note
\ref{enu:Cboe-Global-Markets-VIX}; Wang 2019). Also returns can be
benchmarked against returns from other asset classes outside the crypto
landscape. External indices across asset classes such as stocks, bonds,
commodities and so on could be useful for this purpose (Cizeau et
al., 1997; Leung, Daouk \& Chen 2000; Ratanapakorn \& Sharma 2002;
End-note \ref{enu:Stock-Index}). 

There are further improvements we are planning to the CRI which will
take into account the proportion of assets invested on different chains.
Similar to the basic CRI discussed earlier, the enhanced CRI will
reflect the diversification benefits of amounts invested across multiple
chains (Kashyap 2022). It is relatively straightforward to apply the
CRI to understand macroeconomic concentration risks at a regional,
national or cross national level. This could be done within, or across,
asset classes. The CRI could become an indicator of financial stability
at any desired aggregation level.Considering different aggregation
units, such as countries, can be similar to considering different
networks. 

A historical time series analysis checking how crypto CRI values vary
over time - and their sensitivity to different market events - is
currently being pursued and will be reported in subsequent papers.
Risk management is a vast field filled with numerous challenges. Efforts
are underway to extend the five pillars of crypto risk management
- that we have provided - to the wider risk landscape as well.

\section{\label{sec:Conclusion}Conclusion}

\begin{doublespace}
\noindent We have developed a new metric - dubbed the Concentration
Risk Indicator (CRI) - to measure the concentration risk in any asset
or portfolio. We have discussed how the application of CRI to countries,
regions or any geographic unit can be easily done. We have also shown
that this metric can be applied to risk mitigation in the insurance
sector as well as for any product portfolio mix decisions. The CRI
metric - along with the extensions to portfolios of assets and across
blockchain networks - can be very enlightening for crypto assets and
can be used in traditional financial markets as well. 

\noindent The CRI - when used alongside other well known portfolio
measurement yardsticks - allows us to gauge how portfolios are performing
and to compare them to the wider set of investment opportunities.
We have also given simple formulae - with ways to reduce the amount
of calculations - to calculate Profit and Loss (P\&L), both realized
and unrealized. The metrics we have discussed will be necessary -
once rebalancing and weight calculations function satisfactorily -
to calculate fund performance for regular reporting in terms of risk
and return. 
\end{doublespace}

We have provided a discussion of the fundamental human capital issues
that arise regarding risk management. We pointed out the primary riddle
of risk management, which is that the daily work lives of risk manager
might feel like doing a slow walk on a tight rope. This careful balancing
act between the need for allowing risky trades - to generate profits
- and not taking on too much risk - due to the possibility of incurring
huge losses - can be as daring as the actions of a funambulist. 

We have listed a set of five risk principles - which have been kept
brief purposefully to facilitate easy assimilation within any organization
- that need to be adhered by wealth management firms. It is generally
a troubling scenario wherein someone tries to manage something that
they do not fully understand. Risk, as we know it, requires a lot
more in-depth studies to be performed before we are in a position
to more fully appreciate the etiology of risk. 

Risk management can be done in highly controlled laboratories - or
experimental settings - where the probabilities associated with gains
or losses can be determined very precisely. Such precise measures
to predict outcomes are simply non-existent in the randomness of complex
financial markets or in other real world scenarios. With additional
research and understanding, we will be in a better place to try and
manage risk more throughly. Our present attempts are better referred
to as Risk Mitigation. The need for firms - operating under complicated
situations - to instill a culture that respects - and rewards - both
risk takers and risk mitigaters was discussed. 

Asset selection guidelines, due diligence process, risk management
oversight and the VVV weighting methodology (Kashyap 2022) take care
of monitoring the many other factors that dictate whether an asset
makes a good investment. Continuous innovation - inspired by Sergey
Bubka setting several world records - is the hallmark of any outstanding
portfolio selection, trading and risk management approach. We see
that a combined portfolio of crypto assets can beat BTC in terms risk,
return and CRI. This confirms the intuition that diversification and
better risk management are absolutely necessary for crypto investment
management.With a bedrock of better risk doctrines, blockchain technology
can fulfill its potential to transform all aspects of human interactions.

\section{\label{sec:References}References}
\begin{itemize}
\item Acharya, V. V., \& Richardson, M. P. (Eds.). (2009). Restoring financial
stability: how to repair a failed system (Vol. 542). John Wiley \&
Sons.
\item Akbulaev, N., Mammadov, I., \& Hemdullayeva, M. (2020). Correlation
and Regression Analysis of the Relation between Ethereum Price and
Both Its Volume and Bitcoin Price. The Journal Of Structured Finance,
26(2), 46-56.
\item Alexander, C. (2005). The present and future of financial risk management.
Journal of Financial Econometrics, 3(1), 3-25.
\item Allen, F., Fatás, A., \& di Mauro, B. W. (2022). Was the ICO Boom
just a Sideshow of the Bitcoin and Ether Momentum?. Journal of International
Financial Markets, Institutions and Money, 80, 101637.
\item Ambler, T. (2008). Marketing metrics. In The marketing book (pp. 452-465).
Routledge.
\item Ang, A., \& Bekaert, G. (2004). How regimes affect asset allocation.
Financial Analysts Journal, 60(2), 86-99.
\item Ante, L., Fiedler, I., \& Strehle, E. (2021). The influence of stable-coin
issuances on cryptocurrency markets. Finance Research Letters, 41,
101867.
\item Ante, L., Fiedler, I., Meduna, M. V., \& Steinmetz, F. (2022). Individual
cryptocurrency investors: Evidence from a population survey. International
Journal of Innovation and Technology Management, 19(04), 2250008.
\item Arjoon, S. (2005). Corporate governance: An ethical perspective. Journal
of business ethics, 61, 343-352.
\item Artzner, P., Delbaen, F., Eber, J. M., \& Heath, D. (1999). Coherent
measures of risk. Mathematical finance, 9(3), 203-228.
\item Asad, M., Tabash, M. I., Sheikh, U. A., Al-Muhanadi, M. M., \& Ahmad,
Z. (2020). Gold-oil-exchange rate volatility, Bombay stock exchange
and global financial contagion 2008: Application of NARDL model with
dynamic multipliers for evidences beyond symmetry. Cogent Business
\& Management, 7(1), 1849889.
\item Asimit, A. V., \& Li, J. (2016). Extremes for coherent risk measures.
Insurance: Mathematics and Economics, 71, 332-341.
\item Asmussen, S., \& Steffensen, M. (2020). Risk and insurance. Springer
International Publishing.
\item Austin, R. (1983). The insurance classification controversy. University
of Pennsylvania Law Review, 131(3), 517-583.
\item Baiynd, A. M. (2011). The trading book: A complete solution to mastering
technical systems and trading psychology. McGraw Hill Professional.
\item Bates, D. S. (2012). US stock market crash risk, 1926–2010. Journal
of Financial Economics, 105(2), 229-259.
\item Beck, G. W. (1973). The role of the auditor in modern society: an
empirical appraisal. Accounting and Business Research, 3(10), 117-122.
\item Bendoly, E., Rosenzweig, E. D., \& Stratman, J. K. (2007). Performance
metric portfolios: a framework and empirical analysis. Production
and Operations Management, 16(2), 257-276.
\item Bharati, P., De, U. K., \& Pal, M. (2015, February). A modified diversity
index and its application to crop diversity in Assam, India. In AIP
Conference Proceedings (Vol. 1643, No. 1, pp. 19-29). American Institute
of Physics.
\item Bohr, J., \& Bashir, M. (2014, July). Who uses bitcoin? an exploration
of the bitcoin community. In 2014 Twelfth Annual International Conference
on Privacy, Security and Trust (pp. 94-101). IEEE.
\item Boreiko, D., \& Risteski, D. (2021). Serial and large investors in
initial coin offerings. Small Business Economics, 57, 1053-1071.
\item Brezina, I., Pekár, J., Čičková, Z., \& Reiff, M. (2016). Herfindahl–Hirschman
index level of concentration values modification and analysis of their
change. Central European journal of operations research, 24, 49-72.
\item Brody, R. G., \& Lowe, D. J. (2000). The new role of the internal
auditor: Implications for internal auditor objectivity. International
Journal of Auditing, 4(2), 169-176.
\item Brown, S. J., Fraser, T. L., \& Liang, B. (2008). Hedge fund due diligence:
A source of alpha in a hedge fund portfolio strategy. Available at
SSRN 1016904.
\item Brzęczek, T. (2020). Optimisation of product portfolio sales and their
risk subject to product width and diversity. Review of managerial
science, 14(5), 1009-1027.
\item Chandler, R. A., Edwards, J. R., \& Anderson, M. (1993). Changing
perceptions of the role of the company auditor, 1840–1940. Accounting
and business research, 23(92), 443-459.
\item Chang, L., \& Shi, Y. (2020). Does Bitcoin dominate the price discovery
of the Cryptocurrencies market? A time-varying information share analysis.
Operations Research Letters, 48(5), 641-645.
\item Chen, R. R., \& Fabozzi, F. J. (2010). A risk-based evaluation of
the free-trader option. Quantitative Finance, 10(3), 235-240.
\item Chen, Z., \& Knez, P. J. (1996). Portfolio performance measurement:
Theory and applications. The Review of Financial Studies, 9(2), 511-555.
\item Chen, Y. (2018). Blockchain tokens and the potential democratization
of entrepreneurship and innovation. Business horizons, 61(4), 567-575.
\item Chowdhury, A. (2016). Is Bitcoin the “Paris Hilton” of the currency
world? Or are the early investors onto something that will make them
rich?. The Journal of Investing, 25(1), 64-72.
\item Christoffersen, P. (2011). Elements of financial risk management.
Academic press.
\item Cizeau, P., Liu, Y., Meyer, M., Peng, C. K., \& Stanley, H. E. (1997).
Volatility distribution in the S\&P500 stock index. Physica A: Statistical
Mechanics and its Applications, 245(3-4), 441-445.
\item Cole, R., Johan, S., \& Schweizer, D. (2021). Corporate failures:
Declines, collapses, and scandals. Journal of Corporate Finance, 67,
101872.
\item Colquitt, L. L., Hoyt, R. E., \& Lee, R. B. (1999). Integrated risk
management and the role of the risk manager. Risk Management and Insurance
Review, 2(3), 43-61.
\item Covello, V. T., \& Mumpower, J. (1985). Risk analysis and risk management:
an historical perspective. Risk analysis, 5(2), 103-120.
\item Cowley, A., \& Cummins, J. D. (2005). Securitization of life insurance
assets and liabilities. Journal of Risk and Insurance, 72(2), 193-226.
\item Csikósová, A., Čulková, K., \& Janošková, M. (2016). Evaluation of
quantitative indicators of marketing activities in the banking sector.
Journal of Business Research, 69(11), 5028-5033.
\item Cumming, D. J., Johan, S. A., \& Zhang, Y. (2019). The role of due
diligence in crowdfunding platforms. Journal of Banking \& Finance,
108, 105661.
\item Cummins, J. D., Smith, B. D., Vance, R. N., \& Vanderhel, J. L. (Eds.).
(2013). Risk classification in life insurance (Vol. 1). Springer Science
\& Business Media.
\item Dhaene, J., Laeven, R. J., Vanduffel, S., Darkiewicz, G., \& Goovaerts,
M. J. (2008). Can a coherent risk measure be too subadditive?. Journal
of Risk and Insurance, 75(2), 365-386.
\item Di Pierro, M. (2017). What is the blockchain?. Computing in Science
\& Engineering, 19(5), 92-95.
\item Djolov, G. (2013). The Herfindahl-Hirschman index as a decision guide
to business concentration: A statistical exploration. Journal of Economic
and Social Measurement, 38(3), 201-227.
\item Drobetz, W., Momtaz, P. P., \& Schröder, H. (2019). Investor sentiment
and initial coin offerings. The Journal of Alternative Investments,
21(4), 41-55.
\item Duffie, D., \& Pan, J. (1997). An overview of value at risk. Journal
of derivatives, 4(3), 7-49.
\item Eccles, R. G. (1991). The performance measurement manifesto. Harvard
business review, 69(1), 131-137.
\item Edeling, A., Srinivasan, S., \& Hanssens, D. M. (2021). The marketing–finance
interface: A new integrative review of metrics, methods, and findings
and an agenda for future research. International Journal of Research
in Marketing, 38(4), 857-876.
\item Efing, M., \& Kampkötter, P. (2020, September). Risk Managers in Banks.
In HEC Paris Research Paper No. FIN-2020-1388, Proceedings of Paris
December 2021 Finance Meeting EUROFIDAI-ESSEC.
\item England, C., \& Fratrik, C. (2018). Where to bitcoin?. Journal of
Private Enterprise, 33(1).
\item Fabozzi, F. A., Simonian, J., \& Fabozzi, F. J. (2021). Risk parity:
The democratization of risk in asset allocation. The Journal of Portfolio
Management, 47(5), 41-50.
\item Fahlenbrach, R., \& Frattaroli, M. (2021). ICO investors. Financial
Markets and Portfolio Management, 35(1), 1-59.
\item Fama, E. F., \& French, K. R. (1992). The cross‐section of expected
stock returns. the Journal of Finance, 47(2), 427-465.
\item Fama, E. F., \& French, K. R. (2018). Volatility lessons. Financial
Analysts Journal, 74(3), 42-53.
\item Fassin, Y., \& Gosselin, D. (2011). The collapse of a European bank
in the financial crisis: An analysis from stakeholder and ethical
perspectives. Journal of business ethics, 102, 169-191. 
\item Goetzmann, W., Ingersoll, J., Spiegel, M., \& Welch, I. (2007). Portfolio
performance manipulation and manipulation-proof performance measures.
The Review of Financial Studies, 20(5), 1503-1546.
\item Graham, A., Davey‐Evans, S., \& Toon, I. (2012). The developing role
of the financial controller: evidence from the UK. Journal of Applied
Accounting Research.
\item Grassi, L., Lanfranchi, D., Faes, A., \& Renga, F. M. (2022). Do we
still need financial intermediation? The case of decentralized finance–DeFi.
Qualitative Research in Accounting \& Management.
\item Griffith, S. J. (2015). Corporate governance in an era of compliance.
Wm. \& Mary L. Rev., 57, 2075.
\item Guenther, C., Hienerth, C., \& Riar, F. (2015). The due diligence
of crowdinvestors: thorough evaluation or gut feeling only?. In Academy
of management proceedings (Vol. 2015, No. 1, p. 16862). Briarcliff
Manor, NY 10510: Academy of Management.
\item Gupta, K. (1987). Contemporary auditing. Tata McGraw-Hill Publishing
Company.
\item Harrington, J. L. (2016). Relational database design and implementation.
Morgan Kaufmann.
\item Hong, H., \& Stein, J. C. (2003). Differences of opinion, short-sales
constraints, and market crashes. The Review of Financial Studies,
16(2), 487-525. 
\item Horcher, K. A. (2011). Essentials of financial risk management. John
Wiley \& Sons.
\item Hrisak, D. (1996). The controller as business strategist. Management
Accounting (USA), 78(6), 48-50.
\item Hubbard, D. W. (2020). The failure of risk management: Why it's broken
and how to fix it. John Wiley \& Sons.
\item Hyndman, R. J., \& Fan, Y. (1996). Sample quantiles in statistical
packages. The American Statistician, 50(4), 361-365.
\item Jacobson, D., Brail, G., \& Woods, D. (2012). APIs: A strategy guide.
\textquotedbl{} O'Reilly Media, Inc.\textquotedbl .
\item Jensen, J. R., von Wachter, V., \& Ross, O. (2021). An introduction
to decentralized finance (DeFi). Complex Systems Informatics and Modeling
Quarterly, (26), 46-54.
\item Jorion, P. (1996). Risk\textasciicircum 2: Measuring the risk in
value at risk. Financial analysts journal, 52(6), 47-56. 
\item Jorion, P. (2006). Value at risk: the new benchmark for managing financial
risk. The McGraw-Hill Companies, Inc.. 
\item Jorion, P. (2007). Financial risk manager handbook (Vol. 406). John
Wiley \& Sons. 
\item Kahane, Y. (1977). Determination of the product mix and the business
policy of an insurance company—a portfolio approach. Management Science,
23(10), 1060-1069.
\item Kashyap, R. (2021). Artificial intelligence: A child’s play. Technological
Forecasting and Social Change, 166, 120555.
\item Kashyap, R. (2021b). Do Traders Become Rogues or Do Rogues Become
Traders? The Om of Jerome and the Karma of Kerviel. Corp. \& Bus.
LJ, 2, 88.
\item Kashyap, R. (2022). Bringing Risk Parity To The DeFi Party: A Complete
Solution To The Crypto Asset Management Conundrum. Working Paper.
\item Koenig, D. R. (2006). Aligning Compensation Systems with Risk Management
Objectives. In Risk Management (pp. 703-IX). Academic Press.
\item Kvålseth, T. O. (2018). Relationship between concentration ratio and
Herfindahl-Hirschman index: A re-examination based on majorization
theory. Heliyon, 4(10), e00846.
\item Landier, A., Sraer, D., \& Thesmar, D. (2009). Financial risk management:
When does independence fail?. American Economic Review, 99(2), 454-458.
\item Laine, C. R. (1995). The Herfindahl-Hirschman index: a concentration
measure taking the consumer's point of view. The Antitrust Bulletin,
40(2), 423-432.
\item Le, Y., \& Ieda, H. (2010). Modified Herfindahl–Hirschman index for
measuring the concentration degree of container port systems. Transportation
research record, 2166(1), 44-53.
\item Le Coq, C., \& Paltseva, E. (2009). Measuring the security of external
energy supply in the European Union. Energy policy, 37(11), 4474-4481.
\item Lehmann, D. R. (2004). Metrics for making marketing matter. Journal
of Marketing, 68(4), 73-75.
\item Lee, S., Lee, J., \& Lee, Y. (2022). Dissecting the Terra-LUNA crash:
Evidence from the spillover effect and information flow. Finance Research
Letters, 103590.
\item Leung, M. T., Daouk, H., \& Chen, A. S. (2000). Forecasting stock
indices: a comparison of classification and level estimation models.
International Journal of forecasting, 16(2), 173-190.
\item Lim, C. Y., Woods, M., Humphrey, C., \& Seow, J. L. (2017). The paradoxes
of risk management in the banking sector. The British Accounting Review,
49(1), 75-90. 
\item Lin, T. C. (2016). Compliance, technology, and modern finance. Brook.
J. Corp. Fin. \& Com. L., 11, 159.
\item Linsmeier, T. J., \& Pearson, N. D. (2000). Value at risk. Financial
Analysts Journal, 56(2), 47-67.
\item Liu, W., Sun, Y., Yüksel, S., \& Dinçer, H. (2021). Consensus-based
multidimensional due diligence of fintech-enhanced green energy investment
projects. Financial Innovation, 7, 1-31.
\item Lu, C., Qiao, J., \& Chang, J. (2017). Herfindahl–Hirschman Index
based performance analysis on the convergence development. Cluster
computing, 20, 121-129.
\item Malz, A. M. (2011). Financial risk management: Models, history, and
institutions. John Wiley \& Sons.
\item Marhfor, A. (2016). Portfolio performance measurement: Review of literature
and avenues of future research. American Journal of Industrial and
Business Management, 6(4), 432-438.
\item Matsumoto, A., Merlone, U., \& Szidarovszky, F. (2012). Some notes
on applying the Herfindahl–Hirschman Index. Applied Economics Letters,
19(2), 181-184.
\item McNally, R. C., Durmuşoğlu, S. S., \& Calantone, R. J. (2013). New
product portfolio management decisions: antecedents and consequences.
Journal of Product Innovation Management, 30(2), 245-261.
\item Meng, M., Steinhardt, S., \& Schubert, A. (2018). Application programming
interface documentation: What do software developers want?. Journal
of Technical Writing and Communication, 48(3), 295-330.
\item Merk, O., \& Teodoro, A. (2022). Alternative approaches to measuring
concentration in liner shipping. Maritime Economics \& Logistics,
24(4), 723-746.
\item Mikes, A. (2008). Chief risk officers at crunch time: Compliance champions
or business partners?. Journal of Risk Management in Financial Institutions,
2(1), 7-25.
\item Mohan, V. (2022). Automated market makers and decentralized exchanges:
a DeFi primer. Financial Innovation, 8(1), 20.
\item Murthi, B. P. S., Choi, Y. K., \& Desai, P. (1997). Efficiency of
mutual funds and portfolio performance measurement: A non-parametric
approach. European Journal of Operational Research, 98(2), 408-418.
\item Nakamoto, S. (2008). Bitcoin: A peer-to-peer electronic cash system.
Decentralized business review, 21260.
\item Nietzsche, F. (2019). The use and abuse of history. Dover Publications.
\item Noda, A. (2021). On the evolution of cryptocurrency market efficiency.
Applied Economics Letters, 28(6), 433-439.
\item Noe, T., \& Young, H. P. (2014). The limits to compensation in the
financial sector. Capital failure. Rebuilding trust in financial services,
65-78.
\item Owen, P. D., Ryan, M., \& Weatherston, C. R. (2007). Measuring competitive
balance in professional team sports using the Herfindahl-Hirschman
index. Review of Industrial Organization, 289-302.
\item Parlika, R., \& Pratama, A. (2021, May). Use of the Web API as a basis
for obtaining the latest data on bitcoin prices at 30 exchange places.
In IOP Conference Series: Materials Science and Engineering (Vol.
1125, No. 1, p. 012035). IOP Publishing.
\item Pástor, Ľ. (2000). Portfolio selection and asset pricing models. The
Journal of Finance, 55(1), 179-223.
\item Peng, S., Yang, S., \& Yao, J. (2023). Improving value-at-risk prediction
under model uncertainty. Journal of Financial Econometrics, 21(1),
228-259. 
\item Perez‐Quiros, G., \& Timmermann, A. (2000). Firm size and cyclical
variations in stock returns. The Journal of Finance, 55(3), 1229-1262.
\item Perez, C., Sokolova, K., \& Konate, M. (2020). Digital social capital
and performance of initial coin offerings. Technological forecasting
and social change, 152, 119888. Van Rossum, G., \& Drake Jr, F. L.
(1995). Python tutorial.
\item Pflug, G. C. (2000). Some remarks on the value-at-risk and the conditional
value-at-risk. Probabilistic constrained optimization: Methodology
and applications, 272-281.
\item Power, M. (2004). The risk management of everything. The Journal of
Risk Finance, 5(3), 58-65.
\item Ratanapakorn, O., \& Sharma, S. C. (2002). Interrelationships among
regional stock indices. Review of Financial Economics, 11(2), 91-108.
\item Rauchs, M., \& Hileman, G. (2017). Global cryptocurrency benchmarking
study. Cambridge Centre for Alternative Finance Reports.
\item Rhoades, S. A. (1993). The herfindahl-hirschman index. Fed. Res. Bull.,
79, 188.
\item Rockafellar, R. T., \& Uryasev, S. (2000). Optimization of conditional
value-at-risk. Journal of risk, 2, 21-42. 
\item Rockafellar, R. T., \& Uryasev, S. (2002). Conditional value-at-risk
for general loss distributions. Journal of banking \& finance, 26(7),
1443-1471.
\item Roosenboom, P., van der Kolk, T., \& de Jong, A. (2020). What determines
success in initial coin offerings?. Venture Capital, 22(2), 161-183.
\item Rzepczynski, M. S., \& Black, K. (2021). Alternative Investment Due
Diligence: A Survey on Key Drivers for Manager Selection. The Journal
of Alternative Investments, 24(3), 18-43.
\item Schoonjans, F., De Bacquer, D., \& Schmid, P. (2011). Estimation of
population percentiles. Epidemiology, 22(5), 750-751.
\item Seggie, S. H., Cavusgil, E., \& Phelan, S. E. (2007). Measurement
of return on marketing investment: A conceptual framework and the
future of marketing metrics. Industrial Marketing Management, 36(6),
834-841.
\item Sharma, A., Tiwari, G., \& Rao, K. R. (2020). Identifying mixed use
indicators for including informal settlements as a distinct land use:
Case study of Delhi. Transportation Research Procedia, 48, 1918-1930.
\item Sheikh, U. A., Asad, M., Ahmed, Z., \& Mukhtar, U. (2020). Asymmetrical
relationship between oil prices, gold prices, exchange rate, and stock
prices during global financial crisis 2008: Evidence from Pakistan.
Cogent Economics \& Finance, 8(1), 1757802. 
\item Shi, P., \& Shi, K. (2023). Non-life insurance risk classification
using categorical embedding. North American Actuarial Journal, 27(3),
579-601.
\item Sun, E., \& Baker, L. C. (2015). Concentration in orthopedic markets
was associated with a 7 percent increase in physician fees for total
knee replacements. Health Affairs, 34(6), 916-921.
\item Susilo, Y. O., \& Axhausen, K. W. (2014). Repetitions in individual
daily activity–travel–location patterns: a study using the Herfindahl–Hirschman
Index. Transportation, 41, 995-1011.
\item Travers, F. J. (2011). Investment manager analysis: A comprehensive
guide to portfolio selection, monitoring and optimization. John Wiley
\& Sons.
\item Trucíos, C., Tiwari, A. K., \& Alqahtani, F. (2020). Value-at-risk
and expected shortfall in cryptocurrencies’ portfolio: A vine copula–based
approach. Applied Economics, 52(24), 2580-2593.
\item Truong, T. Q., Zhang, J., Li, Z., \& Wang, L. (2021). Integrated Herfindahl–Hirschman
Index, Compromise Programming, and epsilon-Constraint Method For Multicriteria
Performance-Based Transportation Budget Allocation. Transportation
Research Record, 2675(10), 468-480.
\item Tsai, J. T., Wang, J. L., \& Tzeng, L. Y. (2010). On the optimal product
mix in life insurance companies using conditional value at risk. Insurance:
Mathematics and Economics, 46(1), 235-241.
\item Briola, A., Vidal-Tomás, D., Wang, Y., \& Aste, T. (2023). Anatomy
of a Stablecoin’s failure: The Terra-Luna case. Finance Research Letters,
51, 103358.
\item Uhlig, H. (2022). A Luna-tic Stablecoin Crash (No. w30256). National
Bureau of Economic Research.
\item Urquhart, A. (2022). Under the hood of the ethereum blockchain. Finance
Research Letters, 47, 102628.
\item Van Boxtel, A. A. (2017). Trader Compensation, Risk, and the Banking
Labour Market. Risk, and the Banking Labour Market (March 3, 2017).
\item Van Dijk, M. A. (2011). Is size dead? A review of the size effect
in equity returns. Journal of Banking \& Finance, 35(12), 3263-3274.
\item Van Houwelingen, J. C. (1988). Use and abuse of variance models in
regression. Biometrics, 1073-1081.
\item Veldkamp, L. L. (2005). Slow boom, sudden crash. Journal of Economic
theory, 124(2), 230-257.
\item Wang, H. (2019). VIX and volatility forecasting: A new insight. Physica
A: Statistical Mechanics and its Applications, 533, 121951.
\item Ward, S. (2001). Exploring the role of the corporate risk manager.
Risk management, 7-25.
\item Werner, S. M., Perez, D., Gudgeon, L., Klages-Mundt, A., Harz, D.,
\& Knottenbelt, W. J. (2021). Sok: Decentralized finance (DeFi). arXiv
preprint arXiv:2101.08778.
\item Wolfson, S. N. (2015). Bitcoin: the early market. Journal of Business
\& Economics Research (JBER), 13(4), 201-214.
\item Wu, K., Wheatley, S., \& Sornette, D. (2018). Classification of cryptocurrency
coins and tokens by the dynamics of their market capitalizations.
Royal Society open science, 5(9), 180381.
\item Zetzsche, D. A., Arner, D. W., \& Buckley, R. P. (2020). Decentralized
finance. Journal of Financial Regulation, 6(2), 172-203.
\item Zoni, L., \& Merchant, K. A. (2007). Controller involvement in management:
an empirical study in large Italian corporations. Journal of Accounting
\& Organizational Change, 3(1), 29-43.
\end{itemize}
\begin{doublespace}
\begin{center}
\pagebreak{}
\par\end{center}
\end{doublespace}

\part*{{\LARGE{}\label{part:Appendix-of-Supplementary}Appendix of Supplementary
Material}}

\section{\label{sec:End-notes}Appendix: End-notes and Explanations}
\begin{enumerate}
\item \label{enu:Risk-parity-(or}Risk parity (or risk premia parity) is
an approach to investment management which focuses on allocation of
risk, usually defined as volatility, rather than allocation of capital.
\href{https://en.wikipedia.org/wiki/Risk_parity}{Risk Parity,  Wikipedia Link}
\item \label{enu:Decentralized-finance}Decentralized finance (often stylized
as DeFi) offers financial instruments without relying on intermediaries
such as brokerages, exchanges, or banks by using smart contracts on
a blockchain. \href{https://en.wikipedia.org/wiki/Decentralized_finance}{Decentralized Finance (DeFi), Wikipedia Link}
\item \label{enu:Bitcoin-(BTC)-accounted}Bitcoin (BTC) accounted for 86
percent of the cryptocurrency market, by market capitalization, in
March 2015 and 72 percent in April 2017. Bitcoin’s share of the market
had fallen to just under 48 percent by July 2017. Ethereum (ETH),
the second most popular cryptocurrency, accounted for 22.7 percent
of the market value of cryptocurrencies on July 11, 2017 (Rauchs \&
Hileman 2017; England \& Fratrik 2018). Other cryptocurrencies continue
to increase in size reducing the dominance of BTC and ETH .
\item \label{enu:CoinMarketCap-is-the}
\item \label{enu:Market-capitalization,-sometimes}Market capitalization,
sometimes referred to as market cap, is the total value of a publicly
traded company's outstanding common shares owned by stockholders.
Market capitalization is equal to the market price per common share
multiplied by the number of common shares outstanding. \href{https://en.wikipedia.org/wiki/Market_capitalization}{Market Capitalization,  Wikipedia Link}
\begin{itemize}
\item A similar definition of Market capitalization applies to cryptocurrencies
based on the number of tokens issued by a project and the price of
each token. 
\item \label{enu:Shares-outstanding-are}Shares outstanding are all the
shares of a corporation that have been authorized, issued and purchased
by investors and are held by them. They are distinguished from treasury
shares, which are shares held by the corporation itself, thus representing
no exercisable rights. Shares outstanding and treasury shares together
amount to the number of issued shares.\href{https://en.wikipedia.org/wiki/Shares_outstanding}{Shares Outstanding,  Wikipedia Link}
\item A similar distinction, between total supply versus the circulating
supply of tokens, can be made for cryptocurrencies. 
\end{itemize}
\item \label{enu:Stablecoin}A Stable-coin is a type of cryptocurrency where
the value of the digital asset is supposed to be pegged to a reference
asset, which is either fiat money, exchange-traded commodities (such
as precious metals or industrial metals), or another cryptocurrency.
\href{https://en.wikipedia.org/wiki/Stablecoin}{Stable Coin,  Wikipedia Link}
\item \label{enu:The-Herfindahl-index}The Herfindahl index (also known
as Herfindahl-Hirschman Index, HHI, or sometimes HHI-score) is a measure
of the size of firms in relation to the industry they are in and is
an indicator of the amount of competition among them. \href{https://en.wikipedia.org/wiki/Herfindahl\%E2\%80\%93Hirschman_index}{Herfindahl–Hirschman Index, Wikipedia Link}
\item \label{enu:Cboe-Global-Markets-VIX}Chicago Board Options Exchange
(CBOE) Global Markets revolutionized investing with the creation of
the CBOE Volatility Index® (VIX® Index), the first benchmark index
to measure the market’s expectation of future volatility. The VIX
Index is based on options of the S\&P 500® Index, considered the leading
indicator of the broad U.S. stock market. The VIX Index is recognized
as the world’s premier gauge of U.S. equity market volatility. \href{https://www.cboe.com/tradable_products/vix/}{Chicago Board Options Exchange, VIX Link};
\href{https://en.wikipedia.org/wiki/VIX}{Chicago Board Options Exchange VIX, Wikipedia Link}
\item \label{enu:Coherent-Risk-Measure}In the fields of actuarial science
and financial economics there are a number of ways that risk can be
defined; to clarify the concept theoreticians have described a number
of properties that a risk measure might or might not have. A coherent
risk measure is a function that satisfies properties of monotonicity,
sub-additivity, homogeneity, and translational invariance. \href{https://en.wikipedia.org/wiki/Coherent_risk_measure}{Coherent Risk Measure,  Wikipedia Link}
\item \label{enu:Insurance-is-a}Insurance is a means of protection from
financial loss in which, in exchange for a fee, a party agrees to
compensate another party in the event of a certain loss, damage, or
injury. It is a form of risk management, primarily used to hedge against
the risk of a contingent or uncertain loss. \href{https://en.wikipedia.org/wiki/Insurance}{Insurance,  Wikipedia Link}
\item \label{enu:Stock-Index}In finance, a stock index, or stock market
index, is an index that measures a stock market, or a subset of the
stock market, that helps investors compare current stock price levels
with past prices to calculate market performance. \href{https://en.wikipedia.org/wiki/Stock_market_index}{Stock Market Index,  Wikipedia Link}
\item \label{enu:CoinMarketCap}Coinmarketcap is a website that provides
information and data such as prices, trade volumes, market capitalization
on cryptocurrencies. It was founded in 2013 in New York City by Brandon
Chez. \href{https://en.wikipedia.org/wiki/Coinmarketcap}{Coinmarketcap, Wikipedia Link}
\begin{itemize}
\item CoinMarketCap is the world's most-referenced price-tracking website
for cryptoassets in the rapidly growing cryptocurrency space. Its
mission is to make crypto discoverable and efficient globally by empowering
retail users with unbiased, high quality and accurate information
for drawing their own informed conclusions. https://coinmarketcap.com/about/
\href{https://coinmarketcap.com/about/}{About CoinMarketCap, CoinMarketCap Website Link}
\end{itemize}
\item \label{enu:Structured-Query-Language}Structured Query Language (SQL)
(S-Q-L, sometimes \textquotedbl sequel\textquotedbl{} for historical
reasons) is a domain-specific language used to manage data, especially
in a relational database management system (RDBMS). \href{https://en.wikipedia.org/wiki/Coinmarketcap}{Coinmarketcap, Wikipedia Link}
\item \label{enu:An-application-programming}An application programming
interface (API) is a way for two or more computer programs or components
to communicate with each other. It is a type of software interface,
offering a service to other pieces of software. \href{https://en.wikipedia.org/wiki/API}{API,  Wikipedia Link}
\item \label{enu:Python-is-a}Python is a high-level, general-purpose programming
language. Its design philosophy emphasizes code readability with the
use of significant indentation. It is often described as a \textquotedbl batteries
included\textquotedbl{} language due to its comprehensive standard
library. \href{https://en.wikipedia.org/wiki/Python_(programming_language)}{Python\_(programming\_language),  Wikipedia Link}
\item \label{enu:Crypto-Ranking}A ranking of cryptocurrencies, including
symbols for the various tokens, by market capitalization is available
on the CoinMarketCap website. We are using the data as of May-25-2022,
when the first version of this article was written. \href{https://coinmarketcap.com}{CoinMarketCap Cryptocurrency Ranking,  Website Link}
\begin{doublespace}
\item \label{enu:Artificial-Disciplines}As a first step, we recognize that
one possible categorization of different fields can be done by the
set of questions a particular field attempts to answer. We are the
creators of different disciplines but not the creators of the world
- based on our present state of understanding - in which these fields
need to operate. Hence, the answers to the questions posed by any
domain can come from anywhere or from phenomena studied under a combination
of many other disciplines. This implies that the answers to the questions
posed under the realm of risk management can come from seemingly diverse
subjects such as: physics, biology, psychology, mathematics, chemistry,
marketing, finance, engineering, economics, literature, theater, music
and so on. 
\end{doublespace}
\begin{enumerate}
\begin{doublespace}
\item This suggests that we might be better off identifying ourselves with
problems and solutions, which tacitly confers upon us the title Problem
Solvers, instead of calling ourselves physicists, biologists, psychologists,
mathematicians, engineers, chemists, marketing experts, financiers
,economists, and the like. 
\end{doublespace}
\end{enumerate}
\item \label{enu:Risk-management-is}Risk management is the identification,
evaluation, and prioritization of risks (defined in ISO 31000 as the
effect of uncertainty on objectives) followed by coordinated and economical
application of resources to minimize, monitor, and control the probability
or impact of unfortunate events or to maximize the realization of
opportunities. Risk management, in general, is the study of how to
control risks and balance the possibility of gains; it is the process
of measuring risk and then developing and implementing strategies
to manage that risk. \href{https://en.wikipedia.org/wiki/Risk_management}{Risk Management,  Wikipedia Link}
\item \label{enu:Financial-risk-management}Financial risk management is
the practice of protecting economic value in a firm by managing exposure
to financial risk - principally operational risk, credit risk and
market risk, with more specific variants as listed aside. As for risk
management more generally, financial risk management requires identifying
its sources, measuring it, and the plans to address them. Financial
risk management is the practice of protecting corporate value generally
by using financial instruments to manage exposure to risk, here called
\textquotedbl hedging\textquotedbl ; the focus is particularly on
credit and market risk, and in banks, through regulatory capital,
includes operational risk. \href{https://en.wikipedia.org/wiki/Financial_risk_management}{Financial Risk Management,  Wikipedia Link}
\item \label{enu:A-financial-controller,}A financial controller, or comptroller
(pronounced either the same as controller or as /k\textschwa mp\textprimstress tro\textupsilon l\textschwa r/),
is a management-level position responsible for supervising the quality
of accounting and financial reporting of an organization. A financial
comptroller is a senior-level executive who acts as the head of accounting,
and oversees the preparation of financial reports, such as balance
sheets and income statements. \href{https://en.wikipedia.org/wiki/Comptroller}{Financial Controller,  Wikipedia Link}
\begin{enumerate}
\item Internal control, as defined by accounting and auditing, is a process
for assuring of an organization's objectives in operational effectiveness
and efficiency, reliable financial reporting, and compliance with
laws, regulations and policies. A broad concept, internal control
involves everything that controls risks to an organization. \href{https://en.wikipedia.org/wiki/Internal_control}{Internal Control,  Wikipedia Link}
\end{enumerate}
\item \label{enu:Regulatory-compliance-describes}Regulatory compliance
describes the goal that organizations aspire to achieve in their efforts
to ensure that they are aware of and take steps to comply with relevant
laws, policies, and regulations. \href{https://en.wikipedia.org/wiki/Regulatory_compliance}{Regulatory Compliance,  Wikipedia Link}
\item \label{enu:An-auditor-is}An auditor is a person or a firm appointed
by a company to execute an audit (). \href{\%20https://en.wikipedia.org/wiki/Auditor}{Auditor,  Wikipedia Link}
\begin{enumerate}
\item An audit is an \textquotedbl independent examination of financial
information of any entity, whether profit oriented or not, irrespective
of its size or legal form when such an examination is conducted with
a view to express an opinion thereon.” \href{https://en.wikipedia.org/wiki/Audit}{Audit,  Wikipedia Link}
\item External auditor/ Statutory auditor is an independent firm engaged
by the client subject to the audit, to express an opinion on whether
the company's financial statements are free of material misstatements,
whether due to fraud or error. An external auditor performs an audit,
in accordance with specific laws or rules, of the financial statements
of a company, government entity, other legal entity, or organization,
and is independent of the entity being audited. \href{https://en.wikipedia.org/wiki/External_auditor}{External Auditor,  Wikipedia Link}
\item Internal Auditors are employed by the organizations they audit. Internal
auditing is an independent, objective assurance and consulting activity
designed to add value and improve an organization's operations. It
helps an organization accomplish its objectives by bringing a systematic,
disciplined approach to evaluate and improve the effectiveness of
risk management, control, and governance processes. \href{https://en.wikipedia.org/wiki/Internal_audit}{Internal Auditor,  Wikipedia Link}
\end{enumerate}
\item \label{enu:Percentile}In statistics, a k-th percentile (percentile
score or centile) is a score below which a given percentage k of scores
in its frequency distribution falls (exclusive definition) or a score
at or below which a given percentage falls (inclusive definition).
See: (Hyndman \& Fan 1996; Schoonjans, De Bacquer \& Schmid 2011;
\href{https://en.wikipedia.org/wiki/Percentile}{Percentile,  Wikipedia Link}).
\item \label{enu:Trading-Volume}In capital markets, volume, or trading
volume, is the amount (total number) of a security (or a given set
of securities, or an entire market) that was traded during a given
period of time. (Baiynd 2011; \href{https://en.wikipedia.org/wiki/Volume_(finance)}{Trading Volume,  Wikipedia Link})
\begin{enumerate}
\item The average volume of a security over a longer period of time is the
total amount traded in that period, divided by the length of the period.
Therefore, the unit of measurement for average volume is shares per
unit of time, typically per trading day.
\end{enumerate}
\item \label{enu:Value-at-risk}Value at risk (VaR) is a measure of the
risk of loss for investments. It estimates how much a set of investments
might lose (with a given probability), given normal market conditions,
in a set time period such as a day. \href{https://en.wikipedia.org/wiki/Value_at_risk}{Vaule at Risk, Wikipedia Link}
\item \label{enu:Expected-shortfall-(ES)}Expected shortfall (ES) is a risk
measure—a concept used in the field of financial risk measurement
to evaluate the market risk or credit risk of a portfolio. The \textquotedbl expected
shortfall at q\% level\textquotedbl{} is the expected return on the
portfolio in the worst $q\%$ of cases. ES is an alternative to value
at risk that is more sensitive to the shape of the tail of the loss
distribution. \href{https://en.wikipedia.org/wiki/Expected_shortfall}{Expected Shortfall,  Wikipedia Link}
\item \label{enu:Sergey-Nazarovich-Bubka}Sergey Nazarovich Bubka (born
4 December 1963) is a Ukrainian former pole vaulter. He represented
the Soviet Union until its dissolution in 1991. Sergey has also beaten
his own record 14 times. He was the first pole vaulter to clear 6.0
meters and 6.10 meters. Bubka was twice named Athlete of the Year
by Track \& Field News and in 2012 was one of 24 athletes inducted
as inaugural members of the International Association of Athletics
Federations Hall of Fame. \href{https://en.wikipedia.org/wiki/Sergey_Bubka}{Sergey Bubka, Wikipedia Link} 
\begin{enumerate}
\item Pole vaulting, also known as pole jumping, is a track and field event
in which an athlete uses a long and flexible pole, usually made from
fiberglass or carbon fiber, as an aid to jump over a bar. \href{https://en.wikipedia.org/wiki/Pole_vault}{Pole Vault,  Wikipedia Link} 
\end{enumerate}
\item \label{EN:Net-Asset-Value}Net Asset Value is the net value of an
investment fund's assets less its liabilities, divided by the number
of shares outstanding. \href{https://www.investopedia.com/terms/n/nav.asp}{NAV,  Investopedia Link}
\item \label{enu:Finance-AUM}In finance, assets under management (AUM),
sometimes called fund under management, measures the total market
value of all the financial assets which an individual or financial
institution—such as a mutual fund, venture capital firm, or depository
institution—or a decentralized network protocol controls, typically
on behalf of a client. \href{https://en.wikipedia.org/wiki/Assets_under_management}{Assets Under Management, Wikipedia Link}
\item \label{enu:TVL}In decentralized finance, Total value locked represents
the number of assets that are currently being staked in a specific
protocol.\href{https://coinmarketcap.com/alexandria/glossary/total-value-locked-tvl}{Total Value Locked, CoinMarketCap Link}
\end{enumerate}

\end{document}